\documentclass[prb,aps,nobibnotes,twocolumn]{revtex4}%endfloats
\usepackage{graphicx}%
\usepackage{dcolumn}
\usepackage{amsmath}   
\usepackage{bm}
\usepackage{dcolumn}
\usepackage{longtable}
\usepackage{pifont,color}
\begin{document}

\title{\centering\Large\bf Gaussian excitations model for glass-former dynamics and
                           thermodynamics } 
\author{Dmitry V.\ Matyushov}
\email[E-mail:]{dmitrym@asu.edu.}
\affiliation{
  Department of Chemistry and Biochemistry and Department of Physics and Astronomy, 
  Arizona State University, PO Box 871604, Tempe, AZ 85287-1604}
\author{C.\ A.\ Angell} 
\email[E-mail:]{caangell@asu.edu.}  
\affiliation{
  Department of Chemistry and Biochemistry, 
  Arizona State University, PO Box 871604, Tempe, AZ 85287-1604}
\date{\today}
\begin{abstract}
  We describe a model for the thermodynamics and dynamics of
  glass-forming liquids in terms of excitations from an ideal glass
  state to a Gaussian manifold of configurationally excited states.
  The quantitative fit of this three parameter model to the
  experimental data on excess entropy and heat capacity shows that
  ``fragile'' behavior, indicated by a sharply rising excess heat
  capacity as the glass transition is approached from above, occurs in
  anticipation of a first-order transition -- usually hidden below the
  glass transition -- to a ``strong'' liquid state of low excess
  entropy.  The distinction between fragile and strong behavior of
  glass-formers is traced back to an order of magnitude difference in
  the Gaussian width of their excitation energies. Simple relations
  connect the excess heat capacity to the Gaussian width parameter,
  and the liquid-liquid transition temperature, and strong, testable,
  predictions concerning the distinct properties of energy landscape
  for fragile liquids are made.  The dynamic model relates relaxation
  to a hierarchical sequence of excitation events each involving the
  probability of accumulating sufficient kinetic energy on a separate
  excitable unit. Super-Arrhenius behavior of the relaxation rates,
  and the known correlation of kinetic with thermodynamic fragility,
  both follow from the way the rugged landscape induces fluctuations
  in the partitioning of energy between vibrational and
  configurational manifolds. A relation is derived in which the
  configurational heat capacity, rather than the configurational
  entropy of the Adam Gibbs equation, controls the temperature
  dependence of the relaxation times, and this gives a comparable
  account of the experimental observations.  The familiar coincidence
  of zero mobility and Kauzmann temperatures is obtained as an
  approximate extrapolation of the theoretical equations. The
  comparison of the fits to excess thermodynamic properties of
  laboratory glass-formers, and to configurational thermodynamics from
  simulations, reveals that the major portion of the excitation
  entropy responsible for fragile behavior resides in the
  low-frequency vibrational density of states.  The thermodynamic
  transition predicted for fragile liquids emerges from beneath the
  glass transition in case of laboratory water, and the unusual heat
  capacity behavior observed for this much studied liquid can be
  closely reproduced by the model.
\end{abstract}
\preprint{Submitted to J.\ Chem.\ Phys.\ }
\maketitle

\section{Introduction}
\label{sec:1}
Viscous liquids close to the glass transition are usually
characterized by a broad distributions of (their) relaxation times
(stretched exponential kinetics) and stronger than Arrhenius
(super-Arrhenius) dependence of the relaxation times on
temperature.\cite{Ngai:00,Angell:95} The stretched exponential
kinetics is often represented by Kohlrausch-Williams-Watts (KWW)
relaxation function
\begin{equation}
  \label{eq:0}
  \phi^{\text{KWW}}(t) = \exp\left[-(t/ \tau)^{\beta} \right] ,
\end{equation}
where $\beta \leq 1$ is a stretching exponent and $\phi^{\text{KWW}}(t)$ is a
normalized function representing some relaxing property.  The
temperature dependence of the relaxation time $\tau$ in this equation has
been given many forms,\cite{AngellJAP:00} but is most commonly
represented by the empirical Vogel-Fulcher-Tammann (VFT) law
\begin{equation}
  \label{eq:1}
  \ln(\tau/ \tau_0) = DT_0/(T-T_0) ,
\end{equation}
were $\tau_0$ is a characteristic time for liquid quasi-lattice vibrations.
According to this equation, the relaxation time diverges at
the VFT temperature $T_0$ below the glass transition
temperature $T_g$; $T_0$ is close to $T_g$ for fragile liquids with
super-Arrhenius kinetics and tends to zero for strong liquids with
nearly Arrhenius relaxation.

At sufficiently low temperature, below the range of temperatures
described by the mode-coupling theory,\cite{Goetze:91} a liquid can be
characterized by a set of minima (inherent structures) of the
potential energy landscape divided by Stillinger into ``basins of
attraction''.\cite{Debenedetti:01} The relaxation then occurs by a
sequence of activated transitions\cite{Goldstein:69} between basins or
collections of basins separated by low barriers
(metabasins).\cite{Doliwa:03,Denny:03,Heuer:05} While opinions of its
origin differ, there is much evidence for a dynamic crossover
temperature in fragile liquids, usually corresponding to the critical
temperature of mode coupling theory, where relaxation times are about
$10^{-7}$ s.\cite{Novikov:03} Phenomenological theories, usually
operating below this critical temperature, differ in their attribution
of the driving force responsible for activated events.

The Adam-Gibbs theory\cite{Adam:65} suggests that entropy controls the
activation process. According to this line of
thought,\cite{Gibbs:58,DiMarsio:97,DudowiczReview:06} the divergence
of the relaxation time at $T_0$ is caused by a decrease of the number
of available states related to the configurational entropy $S_c(T)$.
The relaxation time is given by the Adam-Gibbs relation
\begin{equation}
  \label{eq:2}
  \ln(\tau/ \tau_0) = \frac{\Delta}{T S_c(T)}
\end{equation}
which anticipates a growing length-scale $\xi$ attributed to
cooperatively rearranging regions. $\xi$ diverges as $S_c(T)^{-1}$ on
approach to the Kauzmann temperature $T_K$ at which the
configurational entropy
vanishes:\cite{Kauzmann:48,Adam:65,DebenedettiBook:96}
\begin{equation}
  \label{eq:2-1}
  S_c(T_K) = 0 .
\end{equation}

The free volume theory\cite{Cohen:59,Cohen:81} also considers entropy
as the driving force for the glass transition in terms of the volume
available for reorganizing the liquid. This ``free'' volume becomes
increasingly scarce on cooling, eventually leading to a divergent
relaxation time at the VFT temperature. The analysis of $T,P$
viscosity data, however, indicates that it is temperature and not
density that is the primary factor behind the super-Arrhenius
kinetics.\cite{Ferrer:98}

The coefficient $\Delta$ in Eq.\ (\ref{eq:2}) can be related to system
properties within the concept of entropic
droplet.\cite{Xia:00,Xia:01,Lubchenko:04} Following arguments by
Bouchaud and Biroli\cite{Bouchaud:04} and by Lubchenko and
Wolynes,\cite{LubchenkoACP} the probability of creation of a droplet
of size $\xi$ is a competition of the surface, $\propto\xi^2$, and entropic
bulk, $\propto \xi^3$, effects:
\begin{equation}
  \label{eq:3}
  P(\xi)\propto \exp\left( -\sigma(\xi)\xi^2 + S_c \xi^3\right) .
\end{equation}
The probability minimizes at the stationary point of the exponent.
Scaling of the surface tension with the droplet size of the form
$\sigma(\xi)\propto \xi^{-1/2} $ results in the Adam-Gibbs law for the relaxation
time, which, in this concept, corresponds to the time of creating a
mobile region rich in configurational entropy. This picture,
however, does not directly address the question of the origin of
viscous flow,\cite{Goldstein:69} assuming that the mosaic structure of
the dynamically exchanging mobile and immobile regions will have all
the properties necessary to facilitate shear relaxation.  The
divergent length-scale of mosaic regions scales
as\cite{Kirkpatrick:89} $\xi \propto S_c(T)^{-2/3}$ in contrast to
$S_c(T)^{-1}$ scaling in the Adam-Gibbs theory.

The Adam-Gibbs relation can be tested by
calorimetry\cite{Angell:91,Richert:98} or by using relaxation times
and configurational entropies from computer
experiments.\cite{Sastry:01,Voivod:01,Mossa:02,Giovambattista:03,Voivod:04,Gebremichael:05,Saika-Voivod:06}
In the former case, one has to assume that $S_c$ can be approximated
by the excess entropy of a liquid over the crystal $S^{\text{ex}}$,
although this is known to be a poor approximation in many
cases.\cite{Goldstein:76,JohariJCP:00,Angell:02,Speedy:02,JohariJPCB:03}
Nevertheless, the Adam-Gibbs equation is known to work well for both
$S^{\text{ex}}$ and $S_c$, although it does not produce perfectly
straight lines for $\ln(\tau)$ vs $1/TS^{\text{ex}}$, in particular for
fragile liquids. The use of $S^{\text{ex}}$ in the Adam-Gibbs equation
generally fails above the temperature $T_B$ at which the dynamics
change character (e.g.\ bifurcation into $\alpha$ and $\beta$
processes).\cite{Richert:98} The Adam-Gibbs equation with
$S^{\text{ex}}$ used for $S_c$ is often equivalent to the VFT equation
because many glass-formers\cite{Privalko:80,Richert:98} empirically
follow the $1/T$ entropy decay:
\begin{equation}
  \label{eq:4}
  S^{\text{ex}}(T) = S_0\left(1- T_K^{\text{exp}} / T \right) ,
\end{equation}
where $T_K^{\text{exp}}$ is what we will call the
\textit{experimental} Kauzmann temperature obtained by extrapolating
$S^{\text{ex}}(T)$ to zero, in contrast to the \textit{thermodynamic}
Kauzmann temperature $T_K$ defined by Eq.\ (\ref{eq:2-1}). Equation
(\ref{eq:4}) often applies to configurational entropies obtained from
simulations, although bilinear in $1/T$ forms have also been used to
fit the data.\cite{Nave:02,Sciortino:05}

A strong experimental argument in favor of the Adams-Gibbs picture is
the near equality\cite{Angell:97} of the Kauzmann temperature
$T_K^{\text{exp}}$ and the VFT temperature $T_0$.  Further evidence is
provided by the good account it gives of the pressure dependence of
the glass transition
temperature.\cite{Goldstein:63,Angell:76,Angell:79} However, the
theoretical relevance of the concept of nucleation of cooperatively
rearranging regions is not clear.  Given the fact that laboratory
glass-formers and model fluids equilibrated at higher temperatures in
computer experiment approximately follow Eq.\ (\ref{eq:4}), one
wonders to what extent the combination of Eqs.\ (\ref{eq:2}) and
(\ref{eq:4}) is just a successful mathematical relation, reproducing
the VFT law in Eq.\ (\ref{eq:1}), and whether activated events in
supercooled liquids are really driven by the entropy.  The resolution
of this problem is important from a general perspective since the
Adam-Gibbs picture puts supercooled liquids in a unique position
within the more general problem of activated events in condensed
matter. For most cases, including the vast majority of chemical
reactions, excess kinetic energy at a transforming unit or a molecular
mode, and not the entropy, is the driving force which lifts the system
to the top of the activation barrier. This, more traditional, view of
relaxation of supercooled liquids is advocated by models that consider
kinetic energy or enthalpy as the driving force of activated
transitions.\cite{Dyre:06}

In models of activation controlled by the kinetic energy, one
considers excitations to a common high-energy level above the
``top'' of the landscape corresponding to high-temperature
diffusion.\cite{Brawer:84} The energy of the starting point at a basin
minimum is treated as a random variable within trap
models\cite{Monthus:96,Heuer:05} or random-walk
models.\cite{Baessler:87,Arhipov:94,Dyre:95} All such models result in
activated kinetics with a temperature-dependent activation energy.
However, since the probability of transition is finite at each
temperature, no divergent relaxation time appears in these models.

A potential advantage of the energy models is the opportunity to unite
the thermodynamics and dynamics of supercooled liquids within one
conceptual framework, and this is the motivation of the present paper.
In the past, the two-state model by Angell and Rao\cite{Angell:72} led
to configurational entropies that are in quite good agreement with
experimental data.\cite{Moynihan:00} The model suggests that the
thermodynamics of super-cooled liquids can be described by an ensemble
of non-interacting two-state excitations,\cite{Garrahan:03} each
creating an excess of entropy.\cite{Goldstein:72} A recent extension
of the model\cite{DMjcp5:05} considered two Gaussian manifolds of
levels (2G model) instead of two discrete states of the original
two-state model.  This variant of the random energy model,
conceptually related to B{\"a}ssler's random-walk
picture,\cite{Baessler:87,Arhipov:94} allowed an accurate fit of the
laboratory and simulation data for heat capacities and configurational
entropies. Here, we present a somewhat simplified version of this
model that considers excitations from the single-energy level of the
ideal glass to a Gaussian manifold of configurationally excited states
(1G model). The thermodynamic analysis is then further used as a basis
for a dynamic model of configurational excitations.

\begin{figure}[htbp]
  \centering
  \includegraphics*[width=6cm]{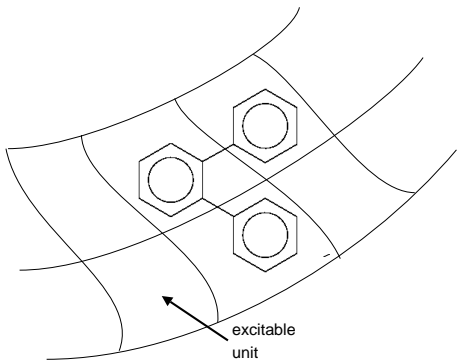}
  \caption{Excitable units of a glass-former in real space.}
  \label{fig:1}
\end{figure}

Our model of dynamics of viscous liquids follows the philosophy of the
energy models of activation in that it considers the probability of
accumulating kinetic energy sufficient to lift a ``structural
element'' (or ``excitable unit'', see Fig.\ \ref{fig:1}) to an energy
level corresponding to the activated state of the high-temperature
relaxation. The kinetic energy supplied by the surroundings of a given
excitable unit is a fluctuating variable with a fluctuation width
related to the ruggedness of the landscape through the configurational
heat capacity. The average of the fluctuations of the kinetic energy
results in a relation for the relaxation time which gives an increase
of the activation energy with decreasing temperature in terms of the
configurational heat capacity, rather than the configurational entropy
of the Adam-Gibbs theory. The new relation gives an account of
experimental dielectric relaxation comparable to the Adam-Gibbs
formula, but more acceptable in several ways to be discussed below.

\section{Thermodynamics of configurational excitations}
\label{sec:2}
\subsection{Formulation of the model}
\label{sec:2-1}
We conceive an excitation\cite{Angell:72,Moynihan:00,DMjcp5:05} to be
a local increase in potential energy that results from a collisional
redistribution of kinetic energy amongst some minimum group of
``rearrangeable units'' of the liquid structure. In this
redistribution a ``unit'' (Fig.\ \ref{fig:1}) that undergoes an
excessively anharmonic vibrational displacement can get trapped by the
motions of neighbors that act to prevent a return of the unit to its
original center of oscillation, such that some kinetic energy will be
lost from the vibrational manifold and stored in the configurational
manifold. It is the constant and repeated exchange of energy between
these manifolds that is the essence of configurational equilibration.
The ability to store potential energy by this mechanism determines the
configurational heat capacity.

\begin{figure}[htbp]
  \centering
  \includegraphics*[width=7cm]{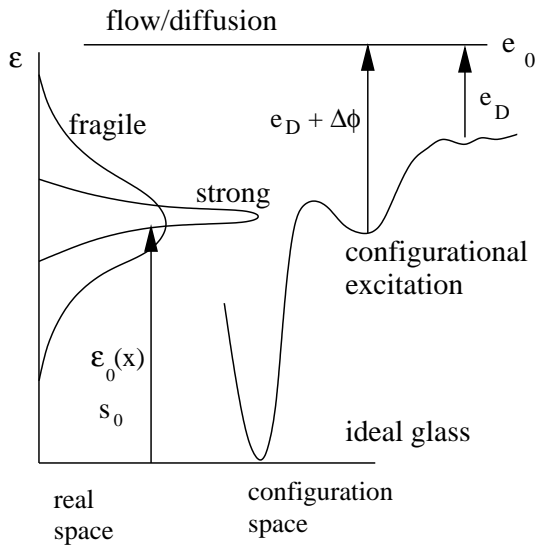}
  \caption{Configurational excitations and flow/relaxation events in
    the 1G model. The thermodynamics of supercooled liquids is
    represented in real space by excitations from a single energy
    level of the ideal glass into a Gaussian manifold of
    configurationally excited states.  Each excitation lifts the
    energy of an excitable unit [Fig.\ \ref{fig:1}] by energy $\epsilon_0(x)$
    [Eq.\ (\ref{eq:8})] and entropy $s_0$. The distribution of these
    states is much broader for fragile liquids than for intermediate
    and strong liquids.  Relaxation is described by the dynamics of
    activated transitions in configuration space from basin minima of
    configurationally excited states to a common energy level $e_0$
    above the top of the energy landscape. The activation barrier
    $e_D+\Delta\phi$ is a sum of the high-temperature activation energy $e_D$
    and the difference $\Delta \phi$ of the average energy of the basin minima from
    the high-temperature plateau. }
  \label{fig:2}
\end{figure}

The mathematical realization of this concept assumes that a liquid can
be divided in real space into excitable units that follow the
statistics of independent entities.  Since these units form a
continuous dense liquid phase and should be interacting, their actual
interactions are represented by mean-field parameters of excitation
energy $\epsilon_0$, excess entropy $s_0$, and excess volume $v_0$. Each
excitable unit is a truly microscopic object as small as a fragment of
a molecule or a single chemical bond of a network glass.  The
low-energy state is identified with the ideal glass, while excitations
associated with molecular motions belong to a Gaussian manifold of
energies with the width $\sigma$ (Fig.\ \ref{fig:2}). The energetic
disorder arises from the local disordering field and packing
restrictions around a given excitable unit.

The real-space model of two-state excitations is next projected onto
Goldstein's configuration space of a liquid at constant
pressure.\cite{Stillinger:88} This configuration space can be
separated into basins of attraction characterized by the minimum
depth $\phi$. We can write the excess Gibbs energy relative to the
energy of the ideal-glass state (superscript ``ex'') as a double sum
over the basin energies and the fraction of excited units $x$ out of
$N$ excitable units:
\begin{equation}
  \label{eq:4-1}
  e^{-g^{\text{ex}}N/T} = \sum_{\phi} e^{-\phi N/T} \sum_{0\leq x\leq 1} e^{s(\phi,x)} ,
\end{equation}
where\cite{Bryngelson:87,Freed:03}
\begin{equation}
  \label{eq:4-3}
  e^{s(\phi,x)} = \frac{N!}{(N-xN)!(xN)!} \left[(Q_v^{e}/Q_v^g)^x P(\phi,x)\right]^{N} .
\end{equation}
Here, $P(\phi,x)$ is the distribution of minimum energies in
configuration space obtained by projecting the excitation energy
$x(\epsilon_0 + Pv_0 + \delta\epsilon)$ on the Gaussian manifold
\begin{equation}
  \label{eq:4-4}
  P(\phi,x) = \int\delta[\phi - x(\epsilon_0 + Pv_0 + \delta\epsilon)] G(\delta\epsilon) d\delta\epsilon ,
\end{equation}
where $G(\delta\epsilon)$ is a Gaussian distribution
\begin{equation}
  \label{eq:4-5}
  G(\delta\epsilon) \propto \exp\left[-(\delta\epsilon)^2/2\sigma^2\right] .
\end{equation}
All energies here and below are in K, entropies and heat capacities
are in units of $k_{\text{B}}$.

The ratio of vibrational-rotational partition functions $Q_v^{g,e}$ in the
ground (superscript ``g'') and excited (superscript ``e'') states can
be absorbed into the excitation entropy
\begin{equation}
  \label{eq:4-2}
  s_0 = \ln\left[Q_v^e/Q_v^g\right] = s_0^{v} + s_0^{c}
\end{equation}
which is composed of the harmonic vibrational contribution, $s_0^{v}$,
and a configurational contribution, $s_0^{c}$.  The vibrational
excitation entropy is related to the excess density of states of
low-frequency vibrational modes near and below the boson
peak:\cite{Angell:03}
\begin{equation}
  \label{eq:4-6}
  s_0^{v} = \sum_{\omega} (g_{\omega}^e - g_{\omega}^g) \ln\left[\omega\right] .
\end{equation}
In Eq.\ (\ref{eq:4-6}), the sum runs over the vibrational frequencies $\omega$
(eigenvalues of the Hessian matrix) with the densities of vibrational states in
the ground and excited states $g_{\omega}^{g,e}$. 

The thermodynamic limit $N \to \infty$ transforms the entropy $s(\phi,x)$ in Eq.\
(\ref{eq:4-1}) into a sum of the ideal mixing entropy, $s_0(x)$, and a
Gaussian term\cite{DMjcp5:05}
\begin{equation}
  \label{eq:4-7}
  s(\phi,x) = s_0(x) - \frac{[\phi - x(\epsilon_0 + Pv_0)]^2}{2x^2\sigma^2},
\end{equation}
where 
\begin{equation}
  \label{eq:4-8}
  s_0(x) = x s_0 - x\ln(x) - (1-x)\ln(1-x). 
\end{equation}
The sum over $x$ in Eq.\ (\ref{eq:4-2}) is determined by its largest
summand at $x=x(\phi)$. One then arrives at the landscape thermodynamics
in which the thermodynamic observables are determined by the excess
free energy function depending on $\phi$ (we omit the dependence on $P$
for brevity)
\begin{equation}
  \label{eq:5}
  \begin{split}
  &g^{\text{ex}}(\phi) = \phi +e_{\text{anh}}^{\text{ex}}(\phi)- T s^{\text{ex}}(\phi),\\ 
  &s^{\text{ex}}(\phi)=s(\phi,x(\phi)),
  \end{split}
\end{equation}
where $e_{\text{anh}}^{\text{ex}}(\phi)$ is the energy related to the
anharmonicity effects not included in the harmonic approximation
(entropy from anharmonicity is small as indicated by computer
simulations\cite{Mossa:02}).  The excess free energy
$g^{\text{ex}}(\phi)$ is composed of the configurational and vibrational
parts
\begin{equation}
  \label{eq:5-2}
  g^{\text{ex}}(\phi) = g_c(\phi) + g_v(\phi),
\end{equation}
where
\begin{equation}
  \label{eq:5-3}
  g_v(\phi) =  e_{\text{anh}}^{\text{ex}}(\phi) - x(\phi) s_0^v T .
\end{equation}

One can alternatively consider the sum over $\phi$ in Eq.\ (\ref{eq:4-1})
that maximizes at the average basin energy $\langle\phi(x)\rangle$.  The (partial)
Gibbs energy can be considered as a function of the population:
\begin{equation}
  \label{eq:5-1}
  g^{\text{ex}}(x) = \langle\phi(x)\rangle  +e_{\text{anh}}^{\text{ex}}(\langle\phi(x)\rangle) - T s(\langle\phi(x)\rangle,x) .
\end{equation}
The minimum of $g^{\text{ex}}(x)$ gives the thermodynamic Gibbs energy
$g^{\text{ex}}$ in Eq.\ (\ref{eq:4-1}).  The formulation in terms of
$g^{\text{ex}}(x)$ is thermodynamically equivalent to the landscape
thermodynamics in terms of $g^{\text{ex}}(\phi)$ since the thermodynamic
Gibbs energy $g^{\text{ex}}$ is achieved at the largest summand in
both $x$ and $\phi$ in the double sum in Eq.\ (\ref{eq:4-1}).  We will,
however, obtain both $g^{\text{ex}}(\phi)$ and $g^{\text{ex}}(x)$ in
order to gain better insight into the physics of the model.

The excitation energy $\epsilon_0$ can, to the first approximation, be
considered as independent of temperature. The situation is quite
different with the Gaussian width $\sigma^2$. The energy of the localized
excited state is randomized by interactions with the thermal motions
of the liquid which are not quenched and therefore affected by
temperature. The fluctuation-dissipation theorem then requires that
$\sigma^2 = 2\lambda T$ scales linearly with temperature,\cite{Landau5,Hansen:03}
as does the mean-square displacement of a classical harmonic
oscillator, $\langle x^2 \rangle \propto T$ . Here, $\lambda$ is the trapping energy or the
energy of stabilization of an excitation by the disorder of the medium
in which it exists.\cite{Baessler:87,Richert:90} Once this real space
Gaussian width is substituted into Eqs.\
(\ref{eq:4-3})--~(\ref{eq:4-5}), it results in an approximately linear
temperature scaling of $\langle(\delta \phi)^2\rangle$ (which is more complex because of a
generally non-parabolic form of $g^{\text{ex}}(\phi)$ and temperature
dependence of the population $x$\cite{DMjcp5:05}).

The notion of the linear ($\propto T$) temperature dependence of $\langle(\delta \phi)^2\rangle$
(which we will verify below) is one of the central components of the
present model.\cite{DMjcp5:05} The \textit{energy} landscape of the
system, i.e.\ energy as a function of $3N$ coordinates of the
molecules making up the liquid, is determined by intermolecular
interactions and is expected to be weakly temperature dependent at
constant volume of the liquid. However, when the manifold of all
possible states in $3N$-space is projected onto one single coordinate
of the basin energy $\phi$, the distribution of $\phi$, given in terms of
the (partial) \textit{free energy} $g^{\text{ex}}(\phi)$, gains
temperature dependence. Not only the first moment of this
distribution, the average basin energy, is temperature dependent, as
indeed described by random-energy models,\cite{Derrida:81} but
essentially all higher moments are temperature-dependent as well. The
random-energy model, originally developed for spin
glasses,\cite{Fischer:99} \textit{assumes} that the width of the
Gaussian distribution of random spin configurations is independent of
temperature. This assumption is well justified for systems with
quenched disorder, but probably not as well for liquids in a
metastable (slow nucleation) equilibrium. Likewise the original
Stillinger-Weber formulation\cite{Stillinger:82} assumed that
temperature affects only the average energy $\langle\phi\rangle$ in the form of
``descending into the landscape'' but not any higher moments of the
distribution.  The simulation evidence on this matter is insufficient
and somewhat controversial. While some simulations of small ensembles
of binary Lennard-Jones (LJ) fluids give clear indication of an
approximately linear dependence of $\langle(\delta \phi)^2\rangle$ on
$T$,\cite{Sastry:98,Heuer:00,Denny:03,Doliwa:03} simulations of larger
systems give virtually constant width.\cite{Sciortino:05} This
distinction is not accidental. The width of the distribution of
inherent structures scales as $1/ \sqrt{N}$ and needs to be measured
on small ensembles.
 
The ideal-glass state should in principle involve randomness
(simulations for network liquids show a finite configurational
entropy at the cutoff energy\cite{Saksaengwijit:04}) and the previous
version of the model\cite{DMjcp5:05} (2G model) assumed Gaussian
distributions for both the ideal glass energies and the excited
configurations with the widths $\sigma_i^2 = 2 k_{\text{B}}T\lambda_i$ ($i=1,2$).
However, as we noted in Ref.\ \onlinecite{DMjcp5:05}, the fit of the
2G model to experimental heat capacities and configurational entropies
resulted in a small and almost constant $\lambda_1\simeq 15-40$ K
for the ideal-glass state.  The application of the model to a more
extensive list of glass-forming liquids performed in this paper has
shown that $\lambda_1$ can be set equal to zero without sacrificing the
quality of the fit.  This will be the model adopted here. This version
of the model, with only one Gaussian manifold for the
configurationally excited states, will be referred to as the 1G model.
The random energy statistics we consider here have much in common with
the model of protein folding proposed by Bryngelson and
Wolynes\cite{Bryngelson:87,Bryngelson:89} and the equations we derive
share features with the earlier cooperative two-state model of
Str\"assler and Kittel\cite{Strassler:65} (which is a forerunner of the
two species non-ideal liquid model of Rapoport,\cite{Rapoport:67} the
``two liquids'' model of Aptekar\cite{Aptekar:79} and Ponyatovsky and
co-workers,\cite{Ponyatovsky:03} and the cooperative defects models of
Granato\cite{Granato:92} and of
Angell-Moynihan\cite{AngellMoynihan:00}).

Now we turn to the thermodynamics of configurational excitations. The
excess Gibbs energy $g^{\text{ex}}(\phi)$ minimizes at the average basin energy
\begin{equation}
  \label{eq:10-2}
  \langle\phi\rangle = x(\epsilon_0+Pv_0) - 2x^2\lambda(1+ \partial e^{\text{ex}}_{\text{anh}} / \partial \phi) .
\end{equation}
In what follows we will neglect the generally unknown derivative 
$ \partial e^{\text{ex}}_{\text{anh}} / \partial \phi$. This approximation is expected to
be accurate at low temperatures close to $T_g$, but will fail at higher
temperatures above the onset temperature at which the system starts to
descent into the energy landscape. In this approximation, the
excited-state population is defined by the self-consistent equation
\begin{equation}
  \label{eq:9}
  x= \left[1 + e^{g_0(x)/T}\right]^{-1} ,
\end{equation}
where the free energy per excitable unit is
\begin{equation}
  \label{eq:10}
  g_0(x) = \epsilon_0(x) + Pv_0 - Ts_0 . 
\end{equation} 
Because of the energetic disorder of the exited states the actual
excitation energy is lowered by twice the trapping energy
$2\lambda$.\cite{WolynesNIST:97,DMjcp5:05} However, only configurationally
excited states can ``solvate'' (stabilize) the excitation, and the
stabilization energy is proportional to the population $x$ of the
excited states.  The effective excitation energy $\epsilon_0(x)$ in Eq.\
(\ref{eq:10}) and Fig.\ \ref{fig:2} then becomes
\begin{equation}
  \label{eq:8}
  \epsilon_0(x) = \epsilon_0 + P v_0 - 2 x \lambda ,
\end{equation}
where the factor of two comes from counting all interactions of a
given unit with the rest of the ensemble of configurational
excitations.

By relating $x$ to the spin variable $\sigma = 2x-1$ Eq.\ (\ref{eq:9}) can
be brought to the form usually considered by models of
ferromagnetism\cite{Ziman:79}
\begin{equation}
  \label{eq:9-1}
 \sigma = \tanh\left[\frac{\lambda\sigma}{2T} - \frac{g_0-\lambda}{2T} \right] ,
\end{equation}
where the excess Gibbs energy of configurational excitations is
\begin{equation}
  \label{eq:9-2}
  g_0 = \epsilon_0 + Pv_0 - Ts_0 .
\end{equation}
At $s_0=0$ and $\lambda=\epsilon_0+Pv_0$ Eq.\ (\ref{eq:9-1}) transforms into the
Weiss formula for spontaneous magnetization\cite{Ziman:79} with $\lambda\sigma/2$
playing the role of the effective field of the magnetic moments.

The self-consistent equation for the population of configurational
excitations [Eq.\ (\ref{eq:9})] bears some similarity with the results of
previous studies minimizing the mean-field free-energy functionals of
liquid-state theories.  The density of the liquid $\rho(\mathbf{r})$
can then be found from the self consistent
equation\cite{Singh:85,Klein:94}
\begin{equation}
  \label{eq:10-1}
    \rho(\mathbf{r}) = q\exp\left(\int c(\mathbf{r}-\mathbf{r}')
                    \rho(\mathbf{r}') d\mathbf{r}' \right), 
\end{equation}
where $q$ is the activity and $c(\mathbf{r})$ is the direct
correlation function.  Solution of this equation predicts a
first-order transition to aperiodic crystal.\cite{Singh:85} MD
simulations, however, indicate\cite{Melcuk:95} that a kinetic
transition happens before the thermodynamic transition is reached, in
qualitative agreement with our analysis of experimental data (see
below).

The combination of the average energy [Eq.\ (\ref{eq:10-2})] and
excess entropy [Eqs.\ (\ref{eq:4-7}) and (\ref{eq:5})] as functions of
population yields $g^{\text{ex}}(x)$ in Eq.\ (\ref{eq:5-1})
\begin{equation}
  \label{eq:10-3}
   \begin{split}
  g^{\text{ex}}(x) &= x g_0 - x^2 \lambda + e_{\text{anh}}^{\text{ex}}(\langle\phi(x)\rangle) \\
            &+ T\left[x\ln x + (1-x) \ln(1-x)\right].
  \end{split}
\end{equation}
Except for the anharmonic correction, this equation has been derived
in many previous
publications.\cite{Strassler:65,Rapoport:67,Aptekar:79,Chamberlin:00,Ponyatovsky:03}
Minimization of $g^{\text{ex}}(x)$ with respect to $x$ (neglecting the derivative
$ \partial e^{\text{ex}}_{\text{anh}} / \partial x$) leads to Eq.\ (\ref{eq:9}).
The parameter $\lambda$ then plays the role of the average energy of
interaction between the configurational excitations.  Consequently,
the quadratic in $x$ term in Eqs.\ (\ref{eq:10-2}) and
(\ref{eq:10-3}), originating from the energy randomness in our model,
is equivalent to direct mean-field (Bragg-Williams\cite{Ziman:79})
interaction between the excited units.  In other words, randomness is
effectively equivalent to attraction when separate units independently
seek the same satisfactory configuration.\cite{WolynesNIST:97} Notice
that derivation of Eq.\ (\ref{eq:10-3}) requires the explicit account
of the linear temperature scaling of the real-space Gaussian width
$\sigma^2$ in Eq.\ (\ref{eq:4-5}). The assumption of a
temperature-independent width would result in a $1/T$ scaling of the
energy term in $g^{\text{ex}}(x)$ quadratic in $x$.

Since the full calculation of the excess thermodynamics of the
supercooled liquid over its crystal is too a complex task even for
phenomenological models we next \textit{assume} that the excess
entropy of configurational excitations over the ideal-glass state,
$s^{\text{ex}}(T)$, gives the excess entropy of the liquid over its
crystal. We will use this excess entropy for the rest of our
thermodynamic analysis warning at this point that this entropy is not 
thermodynamically consistent with the excess Gibbs energies in Eqs.\
(\ref{eq:5}) and (\ref{eq:10-3}), i.e.\ $s^{\text{ex}}(T) \neq -\left(\partial
  g^{\text{ex}}(T)/ \partial T\right)_P$. A thermodynamically consistent $g^{\text{ex}}(T)$
can of course be obtained by temperature integration of $s^{\text{ex}}(T)$.

According to our derivation above, $s^{\text{ex}}(T)$ is the sum of
the configurational entropy $s_c(T)$ and the vibrational entropy
$x(T)s_0^v$:
\begin{equation}
  \label{eq:11-1}
  s^{\text{ex}}(T) =s(\langle\phi(x(T))\rangle,x(T)) = s_c(x(T),T) + x(T) s_0^v,  
\end{equation}
where $\langle \phi(x(T))\rangle$ and $x(T)$ are given by Eqs.\ (\ref{eq:10-2}) and
(\ref{eq:9}), respectively.  The configurational entropy thus
accommodates only that part of $s_0$ which is not related to the
change in the vibrational density of states.  It is given as a sum of
the ideal mixture entropy $s_0^c(x)$ and the fluctuation entropy
$-x^2\lambda/T$ consequent on the shrinking of the Gaussian width with
decreasing temperature:\cite{DMjcp5:05}
\begin{equation}
  \label{eq:11}
    s_c(x,T)  = s_0^c(x) -   x^2 \lambda/T ,
\end{equation}
where
\begin{equation}
  \label{eq:12}
  s_0^c(x) = xs_0^c - x\ln x - (1-x) \ln(1-x) .
\end{equation}

\begin{widetext}
\begin{table*}
  \centering
  \caption{Number of excitable units (``beads'') per mole of a glass-former. The
    last column represents the effective number of classical oscillators in a molecule
    (mole) of a substance at $T_g$. }
  \label{tab:0}
\begin{ruledtabular}
  \begin{tabular}{lccccc}
   Substance & Privalko\cite{Privalko:80} & Takeda\cite{Takeda:99} &
   Moynihan\cite{Moynihan:00} & Stevenson\cite{Stevenson:05} & $C_P^{\text{ex}}(T_g)/3$\\
    \hline
   \textbf{Inorganics} &   &  &  &     &  \\
   Se                 & 1  &1 &1 & 0.9 & 1\\
   ZnCl$_2$             & 1 &   &  & 1.2 & 3\\
   \textbf{Aromatics}  &   &   &  &     &  \\
   Toluene            & 4   & & 1  & 2.7 & 3.2\\
   $o$-terphenyl (OTP) & 9 & & 2 & 3.7 &     \\
   \textbf{Paraffinics} &  & &   &     &     \\
   2-methylpentane    &  6  & & 3 &  3.8 & 3.2 \\
   \textbf{Alcohols}  &    &  &   &    &    \\
   methanol           &  2  &  2  &   & 1.3 & 1.76 \\
   glycerol           &  6  &  6  &  7   &  4.5 & 3.5\\
   \textbf{Hydrates}   &    &  &   &    &    \\
   Ca(NO$_3$)$_2\cdot$4H$_2$O  &  13  &   &   & 7.0 & 11.7 \\
  \end{tabular}
\end{ruledtabular}
\end{table*}
\end{widetext}

\subsection{Application to experimental data}
\label{sec:2-2}
We now proceed to applying the model to laboratory and simulation data
for supercooled molecular liquids.  In the former case, excess entropy
$\Delta s$ of the liquid over the crystal is identified with the excess
entropy over the ideal-glass state $s^{\text{ex}}(T)$ [Eq.\
(\ref{eq:11-1})].  In the case of simulations, conjugate gradient
minimization of simulated trajectories allows sampling of inherent
structures, and equations for the configurational component of the
excess entropy will be used [Eqs.\ (\ref{eq:11}) and
(\ref{eq:12})]. The two sets of equations are formally equivalent when
the overall excitation entropy $s_0$ is used for the excess data and
only its configurational component $s_0^c$ is used for the statistics
of inherent structures.

No progress can be made without recognizing that the molar quantity,
heat capacity, determined in a laboratory measurement contains
contributions from $z$ ``re-arrangeable subunits'' of the molecule
which are the microscopic dynamic elements of the system. This is most
obvious in the case of chain molecule systems like selenium or like
4-methyl nonane\cite{Privalko:80} where each methyl group constitutes
one ``bead'' according to Wunderlich's original
description.\cite{Wunderlich:60} The number of subunits or beads per
molecule is not always evident. In selenium the atom is clearly the
``bead'' but in the glass-former 9-bromo
phenanthrene,\cite{WangAngell:02} the whole C$_{20}$ ``raft'' is rigid
and can only rearrange as a single unit.

Stevenson and Wolynes have suggested an operational approach to
determining $z$, based on the entropy of fusion relative to that of
the simple Lennard Jones (LJ) system.\cite{Stevenson:05} While this
method does not take account of the fact that the LJ fusion entropy is
determined at a temperature where the liquid is enormously more fluid
than the glass-formers at their melting points (entropy per bead 1.68
$k_B$ compared to Wunderlich's 1.36 $k_B$, which leads to gross
overestimates in the case of low fusion entropy network liquids, e.g.\
ZnCl$_2$), it does give a measure that is roughly consistent with
others. These range from the original Wunderlich\cite{Wunderlich:60}
and Privalko\cite{Privalko:80} estimates, through Takeda \textit{et
  al.}  based on structural arguments\cite{Takeda:99} down to Moynihan
and Angell whose estimates were based on best fitting of the excess
entropy to an excitations model.\cite{Moynihan:00} Some comparisons
are provided in Table \ref{tab:0}.

The separation of a molecule into $z$ statistically independent
subunits neglects the finite correlation length of the disordering
field. Indeed, if two units are within the field's correlation length,
they contribute to the statistics of basins as a single unit. The
number of units $z$ is thus an effective parameter necessarily smaller
or equal to the number of conformationally distinct units (cf.\ the
difference between Privalko's conformational and Stevenson's effective
numbers in Table \ref{tab:1}). Because of its effective nature, $z$
can be represented by fractional numbers as was done by Stevenson and
Wolynes (Table \ref{tab:0}). Although this approach provides more
flexibility in fitting the experiment, we will follow here Takeda
\textit{et al.}\cite{Takeda:99} and Moynihan and
Angell\cite{Moynihan:00} and use integral numbers for $z$.  In order
to distinguish between thermodynamic properties referring to excitable
units and molecules (or generally moles), we will use lower-case
letters for the former and upper-case letters for the latter.
Lower-case and upper-case extensive variables are connected through
$z$, e.g.\ for the constant-pressure heat capacity, $C_P = zc_P$.

\begin{widetext}
\begin{table*}[htbp]
  \centering
  \caption{{\label{tab:1}} Best-fit parameters of the 1G model to experimental 
    excess entropies and heat capacities. 
    Also shown are the experimental Kauzmann temperature $T_K^{\text{exp}}$  obtained by extrapolating 
    experimental entropies to zero (Ref.\ 
    \onlinecite{Moynihan:00}) and by using Eq.\ (\ref{eq:16}), experimental VFT temperature
    $T_0$, and the thermodynamic Kauzmann temperature $T_K$ calculated from Eq.\ (\ref{eq:2-1}) using
    the 1G excess entropy from Eqs.\ (\ref{eq:11-1})--(\ref{eq:12}). The temperature of
    liquid-liquid transition $T_{LL}$ is calculated from Eq.\ (\ref{eq:15}), $T_g$ is the
    experimental glass transition temperature. 
    All energies and temperatures are in K, the entropy $s_0$ is in $k_{\text{B}}$ units. }
\begin{ruledtabular}
\begin{tabular}{lccccccccccc}
Substance & $m$\footnotemark[1]$^,$\footnotemark[2] & $z$\footnotemark[3] & $\epsilon_0$ & $\lambda$ & $s_0$ &
           $T_K^{\text{exp}}$ & $T_0$\footnotemark[2] & $T_K^{\text{exp}}$\footnotemark[4] & 
           $T_{LL}$ & $T_K$\footnotemark[5] & $T_g$\\
\hline
   & \multicolumn{11}{c}{Fragile liquids}\\
\hline
Toluene & 105\footnotemark[6] & 1 & 2171 & 1020 & 10.2 & 100  & 96.5\footnotemark[5]  & 100 & 113 & 108 & 117\\
D,L-propene carbonate (PC) & 104 &
                                1 & 2921 & 1383 & 10.7 & 129 &          & 129 & 144 & 123 & 156\\
$o$-terphenyl (OTP)    & 81  & 2 & 3576 & 1686 & 8.3 & 204  & 202.4 & 206 & 228 & 170 & 246\\
2-methyltetrahydrofurane (MTHF) & 65 & 2 & 899 & 414 & 5.9 & 69 & 70 & 70 & 82 & 50 & 91 \\
Salol            & 63  & 3 &  2070 & 988 & 5.6 &  175 & 175 & 176 & 193 & 115 & 220\\
\hline
   & \multicolumn{11}{c}{Intermediate liquids}\\
\hline
3-bromopentane (3BP) & 53 & 4 & 348 & 20 & 2.0 & 84 & 83 &  &          & 75 & 108\\
Glycerol & 53 & 7 & 738 & 5.0 & 1.6  & 137   & 130\footnotemark[7] & & & 59 & 190\\
$n$-propanol (nPOH) & 35 & 2 & 406 & 22 & 2.8 & 72 & 70 & & & 30 & 96\\
\end{tabular}
\end{ruledtabular}  
\footnotetext[1]{Steepness fragility index, Eq.\ (\ref{eq:20}). }
\footnotetext[2]{Taken from Ref.\ \onlinecite{Richert:98} unless indicated otherwise.}
\footnotetext[3]{From Ref.\ \onlinecite{Takeda:99}.}
\footnotetext[4]{Calculated from Eq.\ (\ref{eq:16}). }
\footnotetext[5]{Obtained as the temperature of crossing zero 
                  for the excess entropy calculated in the 1G model.}
\footnotetext[6]{Ref.\ \onlinecite{Doss:97}.}
\footnotetext[7]{Ref.\ \onlinecite{Menon:92}.}
\end{table*}
\end{widetext}

\begin{figure}[htbp]
  \centering
  \includegraphics*[width=6cm]{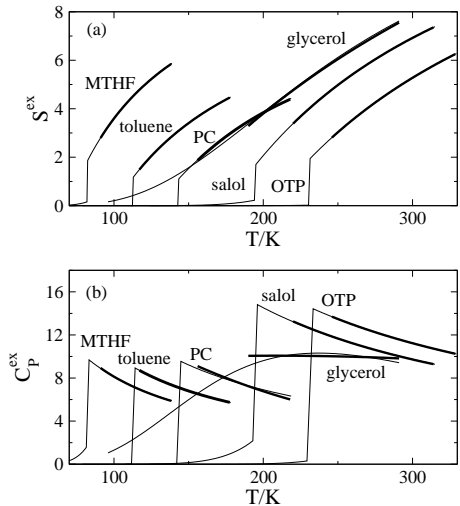}
  \caption{Excess entropy (a) and excess
    heat capacity (b) for some of the supercooled liquids
    listed in Table \ref{tab:1} (per molecule, in units of
    $k_{\text{B}}$). The thin lines refer to fits to the 1G model, the
    thick lines refer to the experimental data.\cite{Moynihan:00} }
  \label{fig:3}
\end{figure}

We will apply the model to constant-pressure data and consider the
excitation energy $\epsilon_0$ as an adjustable parameter. Therefore, for the
rest of our analysis, the term $Pv_0$ in the excess Gibbs energy [Eq.\
(\ref{eq:9-2})] is fused into $\epsilon_0$.  Therefore, given the number of
excitable units per molecule $z$ has been specified, the 1G model
contains three model parameters, $\epsilon_0$, $\lambda$, and $s_0$.  We have
tested the model for a set of glass-formers identified as fragile
(strong super-Arrhenius kinetics, from toluene to salol) and
intermediate (weak super-Arrhenius kinetics, from 3-bromopentane to
$n$-propanol) liquids.  Results shown in Fig.\ \ref{fig:3} and listed
in Table \ref{tab:1} were obtained from a simultaneous fit of the
model to excess heat capacities and entropies from Ref.\
\onlinecite{Moynihan:00}, bound by the constraint $\epsilon_0(x)\geq 0$ [Eq.\
(\ref{eq:8})] required for the mechanical stability of the ideal-glass
state.

The experimental excess entropies are calculated from the
constant-pressure heat capacity $C_P^{\text{ex}}(T)$ and the fusion entropy $\Delta
S_{\text{fus}}$ according to the relation
\begin{equation}
  \label{eq:13}
  S^{\text{ex}}(T) = \Delta S_{\text{fus}} + \int_{T_{\text{fus}}}^{T} C_P^{\text{ex}}(T')(dT'/T'),
\end{equation}
where $T_{\text{fus}}$ is the fusion temperature and
$S^{\text{ex}}(T)=zs^{\text{ex}}(T)$.  The number of excitable units
per molecule (mole) $z$ was taken from Moynihan and
Angell\cite{Moynihan:00} and varied additionally to find the best-fit
integral numbers. Equal quality fits can be obtained in some cases
with different numbers $z$, e.g.\ for toluene $z=1$ and $z=2$ can be
adopted. In that latter case, $z=1$ was taken to maintain the
consistency of parameter values with other fragile liquids (Table
\ref{tab:1}). The choice of $z$ here does not affect our qualitative
conclusions discussed below.

The constant-pressure heat capacity, used to fit the experimental
data, was obtained by the direct differentiation of the excess entropy
constrained by the assumption that the trapping energy $\lambda$ is
independent of temperature. This assumption is supported by
spectroscopic studies showing weak dependence of the Stokes shift
(which is an analog of the trapping energy for electronic transitions)
on temperature.\cite{DMjpca1:06} With this assumption one gets:
\begin{equation}
  \label{eq:14-1}
  c_P^{\text{ex}} = \frac{x^2\lambda}{T} + x(1-x)\frac{\epsilon(x)(\epsilon(x)-2x\lambda)}{T^2(1-x(1-x)(2\lambda/T))} .
\end{equation}
At $\lambda=0$ this equation reduces to Schottky's heat capacity
$c_P^{\text{ex}} = x(1-x)(\epsilon_0/T)^2$, which is further reduced to the
Hirai-Eyring equation,\cite{Hirai:58} proposed on the basis of
transition-state ideas, in the limit $x\ll 1$. For fragile liquids,
$x\simeq 1$, as we show below, and the heat capacity becomes:
\begin{equation}
  \label{eq:16-1}
  c_P^{\text{ex}} = \lambda/T .
\end{equation}
Notice that Eq.\ (\ref{eq:14-1}) anticipates a
Curie-type, $(T-T_c)^{-1}$, divergence of the heat capacity at the
critical temperature defined by the equation $T_c = 2\lambda
x(T_c)(1-x(T_c))$.

\begin{figure}[htbp]
  \centering
  \includegraphics*[width=6cm]{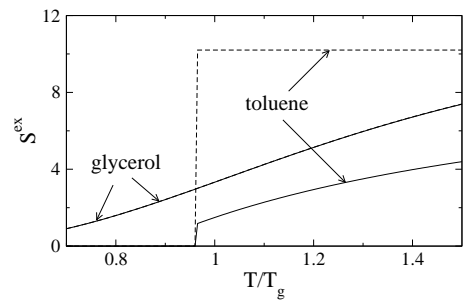}
  \caption{Excess entropy (solid lines) and its ideal-mixture component $zs_0(x)$  
    (dashed lines) [Eq.\ (\ref{eq:4-8})] vs temperature. The dashed and solid lines coincide
    on the scale of the plot for glycerol. }
  \label{fig:4}
\end{figure}

The fragility of a glass-former is often characterized by the
steepness index\cite{Ngai:00,Angell:95}
\begin{equation}
  \label{eq:20}
    m=d\log \tau / d(T_g/T)|_{T=T_g} 
\end{equation}
(listed in Table \ref{tab:1}) or by its thermodynamic
equivalent.\cite{Martinez:01,Richert:06} What appears to be a smooth
transition from fragile salol to intermediate 3-brombenzene according
to the steepness index in fact corresponds to a drastic decrease in
the trapping energy by a factor of about 50. Low values of $\lambda$ turn
out to be characteristic of all intermediate liquids in Table
\ref{tab:1}. The result is a profound change in the relative
importance of the ideal mixing and Gaussian terms in the excess
entropy.

\begin{figure}
  \centering
  \includegraphics*[width=7cm]{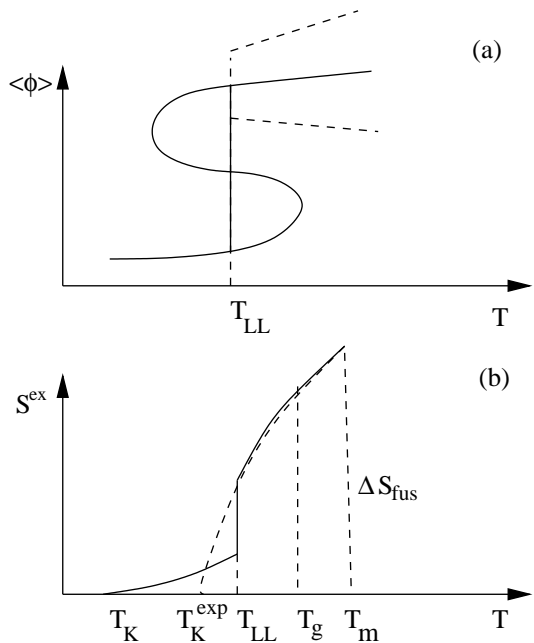}
  \caption{Temperature dependence of the average basin energy (a) and
    excess entropy (b) of fragile glass-formers. The almost flat
    dependence of the average energy on temperature is terminated by
    discontinuous first-order phase transition at temperature
    $T_{LL}$. At this temperature the excess entropy, as predicted by
    the 1G model [solid line in (b)], also shows a discontinuous drop
    to a nonzero value which decays to zero at the thermodynamic
    Kauzmann temperature $T_K$.  The broken line in (b) illustrates
    the experimental interpolation of the high-temperature $1/T$ law
    ending up at the experimental Kauzmann temperature
    $T_K^{\text{exp}}$. The excess entropy is equal to the fusion
    entropy $\Delta S_{\text{fus}}$ at the melting temperature $T_m$. The
    dashed lines in (a) indicate the narrowing of the Gaussian
    distribution of excitation energies with decreasing temperature. }
  \label{fig:5}
\end{figure}

For fragile liquids, the ideal mixing entropy [Eq.\ (\ref{eq:4-8}) or
Eqs.\ (\ref{eq:11-1}) and (\ref{eq:12})] is large and is almost
constant in the whole temperature range of a supercooled liquid $T_g \leq
T \leq T_m$ (toluene in Fig.\ \ref{fig:4}). This is a reflection of the
fact that the population of the high-energy state, driven by the
excitation entropy $s_0$, is close to unity\cite{Chandler:05} and
almost independent of temperature.  The excess ideal mixing entropy is
finally lost in a first-order transition to a low-entropy liquid
state at a temperature $T_{LL}$ which, according to fits to
experimental data, lies between the Kauzmann temperature $T_K$ and
$T_g$ (see Table \ref{tab:1}). The temperature of the equilibrium
phase transition is defined in terms of the model parameters as
\begin{equation}
  \label{eq:15}
  T_{LL} = \frac{\epsilon_0 - \lambda }{s_0} .
\end{equation}

This first-order transition (Fig.\ \ref{fig:5}a) gives rise to a peak
in the heat capacity which is cut off by the kinetic glass transition
at $T_g$.  The observed drop of the heat capacity is related to the
loss of ergodicity which we do not consider here.  The entropy does
not drop to zero at $T_{LL}$ but in all cases becomes small, crossing
zero at the Kauzmann temperature $T_K$ much below the experimental
temperature $T_K^{\text{exp}}$ (Table \ref{tab:1}). A second order
phase transition can be realized at the critical point when the
excitation energy gap is related to the trapping energy by the
following equation\cite{Strassler:65}
\begin{equation}
  \label{eq:15-1}
  \epsilon_0 = \lambda + \lambda s_0/2 . 
\end{equation}
However, when this restriction is imposed on the parameters, the model
fails to fit the experimental data.

The physics is quite different for intermediate liquids for which the
ideal mixture term provides the main part of $s^{\text{ex}}(T)$. The
excess entropy then decreases with temperature due to decreasing
population of the excited state, which cannot stay at a high value
because of a smaller entropy gain $s_0$ compared to fragile liquids
(glycerol in Fig.\ \ref{fig:4}). The entropy smoothly decreases
without a discontinuity related to thermodynamic phase transition (see
Fig.\ \ref{fig:5}b for the illustration of various temperatures used
in the present thermodynamic analysis).

The disappearance of the first-order transition for non-fragile
(strong and intermediate) liquids is related to two critical
parameters, critical temperature $T_c$ and critical excitation
entropy, $s_{0c}$. A first-order phase transition is possible for
temperatures below $T_c$ and entropies higher than
$s_{0c}$:\cite{DMjcp5:05}
\begin{equation}
  \label{eq:21}
  T< T_c,\quad s_0 > s_{0c} ,
\end{equation}
where
\begin{equation}
  \label{eq:22}
  T_c = \lambda / 2, \quad s_{0c}=2 \epsilon_0/ \lambda - 2.
\end{equation}
At least one of two inequalities in Eq.\ (\ref{eq:21}) is violated for
non-fragile liquids as a result of low
excitation entropies $s_0$ and trapping energies $\lambda$.

In parallel to the excess entropy, the average basin energy
$\langle\phi\rangle$ shows qualitatively different temperature behavior for fragile
and intermediate liquids. With temperature decreasing, $\langle\phi\rangle$ from Eq.\
(\ref{eq:10-2}) starts to dip as $1/T$ from a high-temperature plateau
and then inflects into an exponential temperature dependence $\propto
\exp(-\epsilon_0/T)$ for $\epsilon(x)/T\gg s_0$.  Both the $1/T$ decay at relatively
high temperatures\cite{Sastry:98,Buchner:99,Mossa:02,Gebremichael:05}
and exponential decay at low temperatures (for a model network
fluid\cite{Moreno:06}) have been observed in simulations. In our
model, this pattern describes non-fragile liquids.  For
fragile liquids, the excited state population is almost constant, $x\simeq
1$, in the entire range of experimentally accessible temperatures down
to the glass transition.  The average basin energy is constant as
well, $\langle\phi\rangle = \epsilon_0 - 2\lambda $.  For infinitely slow cooling (and rare cases
like triphenyl phosphite (TPP)), the population sharply changes at the
liquid-liquid transition resulting in a discontinuous dip of the
average energy $\langle\phi\rangle$ from its plateau value.

The temperature dependence of the excess entropy of fragile liquids
above $T_g$ is mostly determined by the $1/T$ decay of the second,
fluctuation term in Eq.\ (\ref{eq:11}). The result is the overall
temperature dependence in the form of Eq.\ (\ref{eq:4}) with
$S_0=zs_0$. The experimental Kauzmann temperature is then obtained by
extrapolating Eq.\ (\ref{eq:4}) to zero entropy, which leads, in terms
of the model parameters, to the following relation
\begin{equation}
  \label{eq:16}
  T_K^{\text{exp}} = \frac{\lambda}{s_0}.
\end{equation}
Equation (\ref{eq:16}) holds quite well for the fit parameters in
Table \ref{tab:1}.  In addition, the constant-pressure heat capacity
scales as $1/T$ [Eq.\ (\ref{eq:16-1})].  Therefore, parameter $\lambda$ can
be set by the heat capacity at the glass transition,
\begin{equation}
  \label{eq:18}
  \lambda= T_g c_P^{\text{ex}}(T_g). 
\end{equation}

The fit of experimental data for fragile liquids shows that $\epsilon_0$ is
close to $2\lambda$ and in fact can be put equal to $2\lambda$ without sacrificing
the accuracy of the fit.  As a result, the thermodynamics of fragile
liquids is defined by the following relations often used
empirically\cite{Richert:98,Takeda:99,Xia:01,Gerardin:03,Stevenson:05}
\begin{equation}
  \label{eq:19}
  \begin{split}
  s^{\text{ex}}(T) & = c_P^{\text{ex}}(T_g)\left( T_g/T_K^{\text{exp}} - T_g/T \right), \\
  c^{\text{ex}}_P(T) & = c_P^{\text{ex}}(T_g)(T_g/T) .
  \end{split}
\end{equation}
We note that random energy models, which do not anticipate temperature
variation of the width of basin energy distribution,\cite{Derrida:81}
result in $T^{-2}$ scaling of the heat capacity inconsistent with Eq.\
(\ref{eq:19}).

\begin{figure}
  \centering
  \includegraphics*[width=7cm]{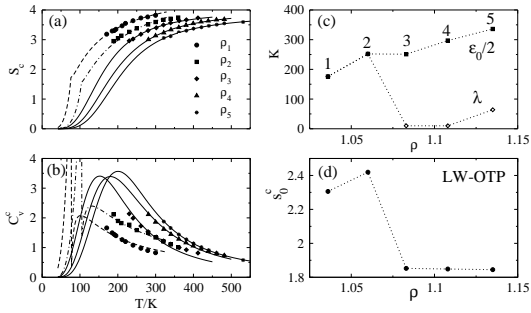}
  \caption{{\label{fig:6}}Configurational entropy (a), configurational
    constant-volume heat capacity (b), and the fitted excitation
    parameters (c,d) for LW $o$-terphenyl ($z=2$ is used as in the
    case of laboratory $o$-terphenyl, Table \ref{tab:1}). The points
    in (a) and (b) are the simulation data from Ref.\
    \onlinecite{Mossa:02} at the densities indicated in (c). The
    dotted lines in (c,d) connect the points. In (c), $\lambda$ is
    indistinguishable from $\epsilon_0/2$ on the scale of the plot for two
    lowest densities. The lines in (a) and (b) are simultaneous fits
    of the simulated configurational entropy and heat capacity to the
    1G model; the dashed and dash-dotted lines are used for two lowest
    densities with high fragility to distinguish them from higher
    densities with intermediate fragility. }
\end{figure}

It is obviously significant for our discussion to compare the results
of fitting the model to laboratory excess entropies with analogous
fits to configurational entropies from simulations. The Lewis and
Wahnstr{\"o}m (LW) model of $o$-terphenyl studied in Refs.\
\onlinecite{Mossa:02} and \onlinecite{Nave:02} provides us with such
an opportunity.  Figure \ref{fig:6} shows the fit of the 1G model to
the \textit{configurational} entropies and heat capacities of LW
$o$-terphenyl at five different densities.\cite{Mossa:02} The number
$z=2$ is maintained equal to the analysis of experimental data to
access the fraction of configurational excitation entropy in $s_0$.
Two main results follow from the fit: (i) the 1G model predicts the
existence of a critical $\lambda$-singularity below $T_g$ for two lowest
densities and its disappearance at higher densities, (ii) the
configurational component of the excitation entropy following from the
fit, $s_0^c\simeq 2$, is significantly lower than $s_0\simeq 8$ obtained for
laboratory $o$-terphenyl (Table \ref{tab:1}). The appearance of the
divergence at lower densities is the result of very close fulfillment
of Eq.\ (\ref{eq:15-1}) for the fitting parameters, which is
remarkable, given that the excitation parameters are freely varied in
the fit. The magnitude of the fitted excitation entropy suggests that
a major portion of $s_0$ arises from a change of the vibrational
density of states,\cite{AngellMoynihan:00,Angell:02,Speedy:02} as we have also learned
from comparing the laboratory and simulation data for water (see
below). Our result is close to Goldstein's estimate of 28\% for the
fraction of configurational component in the excess entropy of
$o$-terphenyl.\cite{Goldstein:76} Goldstein's calculation was based on the
comparison of excess entropies of quenched and annealed glasses at
$T_g$, $\Delta S_{QA}(T_g)$, and at 0 K, $\Delta S_{QA}(0)$. The ratio $\Delta
S_{QA}(0)/\Delta S_{QA}(T_g)$, which is equal to $s_0^c/s_0$ in the 1G
model, gives the configurational fraction of the excess entropy.

The fit of 1G thermodynamic model to experimental data (Table
\ref{tab:1}) shows that the thermodynamic Kauzmann temperature $T_K$
[Eq.\ (\ref{eq:2-1})] is significantly lower than the experimental
Kauzmann temperature $T_K^{\text{exp}}$ [Eq.\ (\ref{eq:4})]. This
implies that the relaxation time in the Adam-Gibbs relation does not
diverge at $T_K^{\text{exp}} \simeq T_0$ and the link between the
thermodynamics and dynamics might be more complex.  The dynamic
extension of the 1G model presented below places the emphasis on the
configurational heat capacity, instead of the configurational entropy,
as the main reason for super-Arrhenius dynamics in fragile liquids.

\section{Dynamics of configurational excitations: enthalpic driving force}
\label{sec:4}
\subsection{Formulation of the model}
\label{sec:4-1}
Here we describe a dynamic model extending the thermodynamic
analysis of Sec.\ \ref{sec:2}. Our development starts with the
assumption common to all energy-trap\cite{Monthus:96,Heuer:05} and
energy-diffusion\cite{Baessler:87,Arhipov:94,Dyre:95} models that
non-reversible events of viscous flow and diffusion\cite{Goldstein:69}
occur by exciting some states within the liquid to a common energy
level $E_0$, which is higher than the top of the energy
landscape\cite{Angell:03,Saksaengwijit:06} (Fig.\ \ref{fig:2}). These
excitations occur by absorbing kinetic energy by an excitable unit
(bead) from the surrounding liquid.

We will next assume that only unjammed configurationally excited units
will participate in activated events. Even when a sufficient amount of
energy has been accumulated at a given unit, relaxation event may
require facilitation from other units.  This is particularly clear in
a case of a molecule composed of $z$ units (beads). One could imagine
that e.g. translational relaxation of such molecule would require
excitation of all $z$ units, although relaxation of conformationally
flexible molecules can also proceed in a diffusive way, as a sequence of
low-amplitude motions of consecutive units. Because of the assumed low
amplitude of the motions involved, we will not distinguish between
pairs of units within the molecule and pairs of units belonging to
different molecules.

\begin{figure}[htbp]
  \centering
  \includegraphics*[width=7cm]{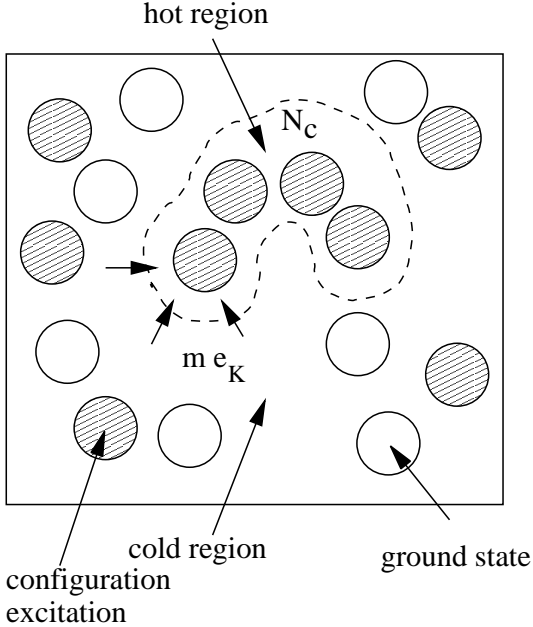}
  \caption{Hot region of $N_c$ configurationally excited units
    (hatched circles) with kinetic energy accumulated by transfer of
    $m$ ``quanta'' of the kinetic energy $e_K$ from the surrounding
    liquid.  }
  \label{fig:7}
\end{figure}

Each low-amplitude step is terminated by a transformation of the
accumulated kinetic energy into a small structural change with a
higher potential energy and thus higher fictive temperature. A single
relaxation event requires a sequence of such locking steps by the end
of which there appears a ``hot'' (in fictive temperature) region
within the liquid which can then subsequently relax to a new
configuration (Fig.\ \ref{fig:7}). In this scheme the overall dynamics
become hierarchical in character, the initial step being the most
probable and hence the shortest in time. This picture is an
integration of energy spikes seen in simulations of Heuer and
co-workers\cite{Buchner:99,Saksaengwijit:06} with dynamical hierarchy
of Palmer \textit{et al.}\cite{Palmer:84} (see below).

The kinetic energy becomes increasingly scarce at low temperatures.
The creation of a hot region will occur by pulling the kinetic energy
from a growing number of neighboring molecules, leading to the
creation of a hot island in a sea of kinetically frozen molecules (on
the time scale of heat transport).  This picture bears some similarity
to the entropy-rich droplet enveloped by the entropy-frozen
environment described by Lubchenko and Wolynes\cite{LubchenkoACP} and
by Bouchaud and Biroli.\cite{Bouchaud:04} However, the notion of
relaxation proceeding by occurrence of hot regions does not anticipate
static, thermodynamically stable structures of the mosaic
picture\cite{Cavagna:06} and instead corresponds to the idea of
dynamic heterogeneity i.e.\ the existence of regions of markedly
different mobility.\cite{Ediger:00,Richert:02} Computer simulations
generally support this
view.\cite{Donati:98,Giovambattista:03,WidmerCooper:04} Within this
general umbrella, dynamic heterogeneity can be treated in two distinct
ways: as a static distribution of relaxation times in regions of
varying mobility\cite{Vilgis:94,Xia:01,Richert:02} or in a dynamic
fashion as facilitated kinetics of transfer of excitations from mobile
particles to their neighbors.\cite{Garrahan:03} Our formulation will
follow this latter pathway using the kinetic scheme of hierarchical
relaxation events advanced by Palmer \textit{et al.}\cite{Palmer:84}

According to the hierarchically facilitated dynamics, step $n+1$
happens only when a configuration at step $n$ is reached to facilitate
the next move.\cite{Palmer:84} This idea has been instrumental in
establishing the conceptual basis for kinetically constrained
models\cite{Ritort:03} and, physically, leads to dynamical
heterogeneity when dynamically cooperative regions in the glass are
created by a sequence of constrained motions of fastest molecules in
the ensemble.\cite{Garrahan:03} In our model, step $n$ is reached when
$n$ excitable units within a hot island have ``blinked'' into the
excited state with the probability
\begin{equation}
  \label{eq:31}
    P_n=(\tau_0/\tau_1)^n=\exp[\mu n],  
\end{equation} 
where $\tau_1$ is the average waiting time for a single unit and $\tau_0$ is
the same as in Eq.\ (\ref{eq:1}). As a result of a sequence of
correlated steps, each resulting in excitation of $n$ units out of
$N_c$ units in the hot island, the waiting time of level $n$
becomes\cite{Palmer:84}
\begin{equation}
  \label{eq:32}
  \tau_n = \tau_0\exp[\mu N_n]
\end{equation}
with $N_n = n(n-1)/2$. We can now follow Brey and Prados\cite{Brey:01}
to obtain the normalized relaxation function:
\begin{equation}
  \label{eq:33}
  \phi(t) = \zeta^{-1}\left[E_1(te^{-\zeta}/ \tau_0 ) - E_1(t/ \tau_0) \right],
\end{equation}
where $E_1(z)$ is the exponential integral function,
\begin{equation}
  \label{eq:34}
    \zeta= \mu N_{\text{max}}  ,
\end{equation}
and 
\begin{equation}
  \label{eq:34-1}
    N_{\text{max}}=N_c(N_c-1)/2.
\end{equation}

\begin{figure}[htbp]
  \centering
  \includegraphics*[width=6cm]{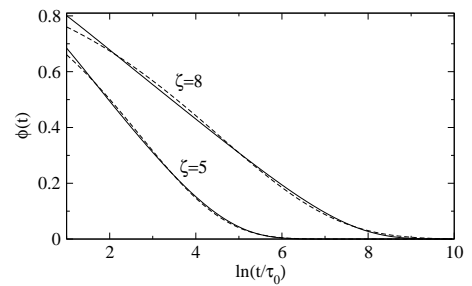}
  \caption{$\phi(t)$ from Eq.\ (\ref{eq:33}) (solid lines) and the
    closest corresponding KWW relaxation function [dashed lines, Eq.\
    (\ref{eq:0})]. The values of $\zeta$ used to plot $\phi(t)$ are indicated
    in the plot.  }
  \label{fig:8}
\end{figure}

For most practical purposes the relaxation function in Eq.\
(\ref{eq:33}) is indistinguishable (Fig.\ \ref{fig:8}) from the KWW
function in Eq.\ (\ref{eq:0}).  In the intermediate range of times,
$1\ll t/ \tau_0 \ll e^{\zeta}$, which can be very broad since $\zeta\gg 1$ for real
systems, $\phi(t)$ follows the logarithmic decay, $\phi(t) \simeq 1 - \ln(t/
\tau_0)$, common for biopolymers.\cite{Abrahams:05} The average
relaxation time $\tau = \int_0^{\infty} t \phi(t) dt$ from Eq.\ (\ref{eq:33}) is
\begin{equation}
  \label{eq:36}
  \tau = (\tau_0/ \zeta) \left(e^{\zeta}-1\right)\simeq \tau_0 e^{\zeta} .
\end{equation}
Notice that the KWW function is more flexible than Eq.\ (\ref{eq:33})
because it involves two free parameters, $\tau/ \tau_0$ and $\beta$, in contrast
to the single parameter $\zeta$ in Eq.\ (\ref{eq:33}). More complex
facilitation rules than the ones used here will provide additional
parameters and a possibility of realizing the KWW relaxation
function\cite{Palmer:84} which has the advantage of allowing an
approximate correlation between the stretch exponent and
fragility.\cite{Boehmer:94}

\begin{figure}
  \centering
  \includegraphics*[width=7cm]{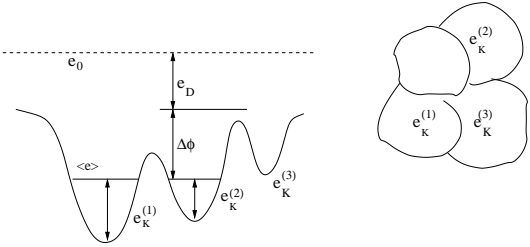}
  \caption{Activation barrier from the average energy $\langle e\rangle$ to the
    energy level $e_0$ above the top of the landscape and fluctuations
    of the kinetic energy $e_K$ in a rugged landscape of fragile
    liquids.  The activation barrier $e_D$ refers to relaxation in a
    high-temperature liquid, $\Delta\phi=\langle \phi(T)\rangle - \langle \phi(\infty)\rangle$ is the change
    of the average minimum energy from the high-temperature plateau.
    The real space regions on the right illustrate fluctuations in the
    kinetic energy of the molecules due to variations of the basin
    depths. }
  \label{fig:9}
\end{figure}

Each elementary step within the hierarchical sequence requires
overcoming the activation barrier between the average energy
of the basin minimum $\langle\phi\rangle$ and the common energy level $E_0=ze_0$ (Fig.\ \ref{fig:9}):
\begin{equation}
  \label{eq:23}
  E_D(x) = E_D - z \Delta \phi(x),
\end{equation}
where $E_D$ is the activation barrier per molecule (mole) associated
with the activated relaxation in the high-temperature
liquid\cite{Frenkel:55} and
\begin{equation}
  \label{eq:24}
  \Delta \phi(x) = (x-1)\epsilon_0 - 2(x^2-1)\lambda 
\end{equation}
is the drop of the minimum energy from the high-temperature plateau
below the onset temperature. 

One can next assume that the kinetic energy necessary for activation
is distributed in ``quanta'' of thermal kinetic energy $E_K=ze_K$
throughout the liquid, where $E_K$ and $e_K$ refer to a molecule
(mole) and excitable unit, respectively. This physical picture seems
appropriate for describing activated events in disordered materials
since, according to theories of heat conductivity, the quasi-lattice
vibrations in glasses are more appropriately described (in the
temperature range above ca.\ 30 K) as quasi-localized vibrations
rather than wave-like motions.\cite{Cahill:88} The momentum exchange
(heat transport) occurs by diffusional transport of vibrational energy
by these quasi-localized modes in the vicinity of the boson
peak.\cite{Sheng:94,Schirmacher:98} Activation of one unit then
requires accumulating $m= E_D(x)/E_K$ ``quanta'' of kinetic energy,
provided that that unit is not jammed being in the configurationally
excited state.  Therefore, the probability of activating one unit is
equal to the probability of absorbing $m$ quanta of kinetic energy out
of a manifold of $xN$ configurational excitations uniformly
distributed over $N$ units (beads) in the liquid.  This type of
problem is considered in the theory of unimolecular bond dissociation.
The solution by Kassel\cite{Forst:03} gives the probability of
combining at least $m$ quanta of energy at one bond:
\begin{equation}
  \label{eq:25}
  P_{\geq m} = \frac{(N+xN - m-1)!(xN)!}{(xN-m)! (N+xN-m)!} .
\end{equation}

For condensed-phase problems one takes the thermodynamic limit in the above
equation, $N \to \infty$, with the result
\begin{equation}
  \label{eq:26}
  P_{\geq m} = (1 + 1/x)^{-m} = \exp\left(-E(x)/E_K \right),
\end{equation}
where
\begin{equation}
  \label{eq:28}
  E(x) = E_D(x) \ln(1+1/x),
\end{equation}
and $E_D(x)$ is given by Eq.\ (\ref{eq:23}). 

Because of similar combinatorial rules, not surprisingly,
Eq.\ (\ref{eq:26}) is analogous to the equation for the probability of 
finding a hole with the volume exceeding some critical volume $v^*$
\begin{equation}
  \label{eq:28-1}
  P(v^*)=\exp\left(-\gamma v^*/v_f \right),
\end{equation}
where $v_f$ is the free volume and $\gamma$ is a numerical coefficient.
This equation is the key result of free-volume models of diffusion and
relaxation in glass-formers.\cite{Cohen:59,Cohen:81} The ideal glass
transition is predicted to occur because the liquid is supposed to run
out of free volume at a finite temperature. In contrast to that, the
kinetic energy in Eq.\ (\ref{eq:26}) is a fluctuating variable which
can approach zero for some basins in the distribution sampled at
temperature $T$ ($e_K^{(3)}$ in Fig.\ \ref{fig:9}), but whose average
value is proportional to $T$.

The combinatorial arguments of the Kassel model envision the system as
a microcanonical ensemble characterized by the average energy $\langle e \rangle $
uniformly distributed over the sample (Fig.\ \ref{fig:9}).  In order
to apply these combinatorial rules to a macroscopic liquid, we need to
use the microcanonical ensemble characterized by the average energy $\langle
e\rangle$ [analogously to the use of microcanonical ensemble to calculate
the excess entropy in Eqs.\ (\ref{eq:4-3}) and (\ref{eq:4-7})]. Since
energy is an extensive variable, this description is equivalent in the
thermodynamic limit to the canonical one in the sense that the
fluctuations of the total energy can be neglected.

Each unit undergoing excitation to the common level $e_0$ by
collecting kinetic energy from the surrounding molecules will find
itself in a local disordering field characteristic of a particular
basin of the rugged energy landscape (Fig.\ \ref{fig:9}).  The kinetic
energy available to the surrounding molecules to excite a given
molecule will be a fluctuating variable producing disorder of $e_K$
which is quenched on the time-scale of momentum relaxation. The
relaxation time of a single molecule $\tau_1$ then needs to be averaged
over possible realizations of the kinetic energy
\begin{equation}
  \label{eq:27}
   \tau_1 = \tau_0e^{\mu} = \tau_0 \int \exp[E(x)/(ze_K)] P(e_K) de_K,
\end{equation}
where $P(e_K)$ is the distribution of the kinetic energy. 

For the energy landscape characteristic of strong and intermediate
liquids the canonical narrow distribution of basin energies projects
itself, in the microcanonical ensemble, into a narrow distribution of
kinetic energy $e_K$. The distribution function $P(e_K)$ in Eq.\
(\ref{eq:27}) can be replaced by a delta function.  Assuming that the
average kinetic energy is equal to $E_K=(3/2)T$ (translations for
diffusion and viscous flow and rotations for dielectric relaxation),
and using Eq.\ (\ref{eq:28}) for $E(x)$ in Eq.\ (\ref{eq:27}) 
one gets for the average relaxation time
\begin{equation}
  \label{eq:37}
  \ln(\tau/ \tau_0)=  (2/3T)(E_D-z\Delta\phi(x)) \ln\left[2 + e^{-s_0 + \epsilon(x)/T} \right] . 
\end{equation}
The relaxation is Arrhenius at high temperatures ($\epsilon(x)/T\ll s_0$ and $z
\Delta\phi \ll E_D$). Two things happen when the temperature
is lowered.  First, the energy gap between the average minimum energy
$\langle\phi_m\rangle$ and the energy level $e_0$ starts to increase\cite{Dyre:95} as
$a + b/T$ since $\Delta \phi(x)$ scales as $1/T$ right below the onset
temperature.\cite{Sastry:01}  Second, at $\epsilon(x)/T\gg s_0$ the logarithmic term in Eq.\
(\ref{eq:37}) generates a $1/T$ factor.  Overall, the temperature law
at these low temperatures becomes
\begin{equation}
 \label{eq:38}
  \ln(\tau/ \tau_0)= E_1/T^2 + E_2/T^3,
\end{equation}
which is a linear combination of the B\"assler\cite{Richert:90} and
Litovitz\cite{Litovitz:52} temperature laws.

\begin{figure}[htbp]
  \centering
  \includegraphics*[width=6cm]{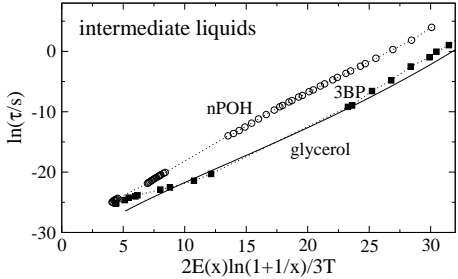}
  \caption{Test of Eq.\ (\ref{eq:37}) using dielectric relaxation data
    for intermediate liquids listed in Table \ref{tab:1}. The
    high-temperature activation energy $E_D$ in Eq.\ (\ref{eq:37}) is
    equal to $18.4T_g$ (glycerol), $22.6T_g$ (nPOH), and $16.0T_g$
    (3BP).\cite{Novikov:05} $E_D$ for 3BP was obtained from its
    empirical connection to the steepness index $m$ given in Ref.\
    \onlinecite{Novikov:05}.  The excited-state population $x(T)$ and
    $\Delta\phi(T)$ in Eq.\ (\ref{eq:37}) are calculated from the 1G model
    with the liquid parameters listed in Table \ref{tab:1}.  The
    slopes of the linear regressions are 1.0 (glycerol), 1.1 (nPOH),
    and 1.0 (3BP), and the intercept gives the physical value of
    $10^{-13}$ s. In case of glycerol, the average kinetic energy of
    $3T$ has been adopted. }
  \label{fig:10}
\end{figure}

This sort of non-Arrhenius kinetics is capable of describing the relaxation of
intermediate liquids studied in Sec.\ \ref{sec:2}. Figure \ref{fig:10}
shows experimental dielectric relaxation times\cite{Richert:98} vs the
rhs of Eq.\ (\ref{eq:37}) calculated from the 1G model. The parameters
affecting population $x$ are taken from our thermodynamic analysis
summarized in Table \ref{tab:1}. Experimental high-temperature
activation energies $E_D$ are from Ref.\ \onlinecite{Novikov:05}.  The
analysis yields straight line with the slope 1.1 for $n$-propanol and a
less clear linear trend for 3BP with the linear regression slope of
1.0. For glycerol, the slope of 1.0 is obtained by assuming the
average kinetic energy equal to $3T$, which might reflect the
participation of both rotational and translational degrees of freedom
in the relaxation process. We need to note that vibrational heat
capacity of many glass-forming liquids does not reach the Dulong and
Petit limit in its low-temperature portion. Therefore, the average
kinetic energy can fall below $(3/2)T$ also becoming a non-linear
function of temperature.  Accounting for this effect, which is not
considered here, will make the temperature dependence in Eq.\
(\ref{eq:37}) even more complex.

Equation (\ref{eq:37}) is insufficient to account for super-Arrhenius
behavior of fragile liquids from Table \ref{tab:1}. This is because
fluctuations of the kinetic energy now need to be taken into account [Eq.\
(\ref{eq:27})].  $E_K$ can be connected to the fluctuation of the
energy of inherent structures by assuming that the average energy is $\langle E \rangle = (3/2)T +
z \langle\phi\rangle$. The instantaneous kinetic energy is then
\begin{equation}
  \label{eq:40}
  E_K = (3/2)T - z \delta\phi ,
\end{equation}
where $\delta\phi = \phi - \langle\phi\rangle$. The distribution of $\delta \phi $ is Gaussian,
\begin{equation}
  \label{eq:41}
  P(\delta\phi) = \left(2\pi c_P^c T^2 \right)^{-1/2} \exp \left[-\frac{(\delta\phi)^2}{2c_P^c T^2}\right] .
\end{equation}
This follows from the bilinear expansion of the enumeration function
$s^c(\phi)$ in $\delta \phi $ and the use of two thermodynamic identities\cite{Landau5}
\begin{equation}
  \label{eq:6}
   \begin{split}
   &\left( \partial s^c(\phi)/ \partial\phi\right)_{N,P} = T^{-1}, \\
   &\left( \partial^2 s^c(\phi)/ \partial\phi^2\right)_{N,P} = -\left(c_P^c T^2\right)^{-1} , 
   \end{split}
\end{equation}
where $c_P^c$ is the configurational heat capacity of the excitable
unit [in contrast to the excess heat capacity in Eq.\
(\ref{eq:14-1})]. Notice regarding Eq.\ (\ref{eq:41}) that Boltzmann
statistics of the basin fluctuations follow from the hyperbolic,
$c_P^c\propto 1/T$ scaling of the configurational heat capacity, while
$1/T^2$ scaling of the random energy models\cite{Derrida:81} leads to
the temperature-independent statistics.

The average waiting time of one-unit excitation then incorporates the thermodynamic
quantity, configurational heat capacity, into the kinetics through $P(\delta\phi)$ 
in the integral:
\begin{equation}
  \label{eq:42}
  \tau_1 / \tau_0 = \int_{-\infty}^{3T/2z} \exp\left[\frac{E(x)}{3T/2 - z \delta\phi }\right] P(\delta\phi) d\delta\phi  . 
\end{equation}
A simple estimate of the integral in Eq.\ (\ref{eq:42}) can be
obtained by linearly expanding the exponent in $\delta\phi$, integrating over
$\delta\phi$, and reverting to the fractional form. Combining
Eqs.\ (\ref{eq:34}), (\ref{eq:36}), and (\ref{eq:42}), the average
relaxation time of the hot region is then obtained as
\begin{equation}
   \label{eq:43}
   \ln\left(\tau/ \tau_0\right) = \frac{2N_{\text{max}}E(x)}{3T - (4E(x)/9)zC_P^c(T)} ,
\end{equation}
where $N_{\text{max}}$ is given by Eq.\ (\ref{eq:34-1}) and the
molecular (molar) configurational heat capacity $C_P^c(T)$ appears in
the denominator.  For fragile liquids, $E(x)$ is nearly independent of
temperature down to the liquid-liquid transition point ($x\simeq 1$, see
Fig.\ \ref{fig:5}a) and can be considered as constant. This
consideration yields the following equation for the relaxation time
 \begin{equation}
   \label{eq:44}
   \ln\left(\tau/ \tau_0\right) = \frac{D T'}{T - T' C_P^c(T)} ,
\end{equation}
where the constant $D$ is
\begin{equation}
  \label{eq:45}
  D = (9/4z) N_c(N_c-1) 
\end{equation}
and the temperature $T'$ is described further below.
 
The parameters $D$ and $T'$ in Eq.\ (\ref{eq:44}) can be considered as
empirical fitting quantities for the sake of interpreting the
experiment. According to Eq.\ (\ref{eq:16-1}), the excess and
configurational heat capacities of fragile liquids are equal to each
other for fragile liquids.  The $1/T$ scaling of the configurational
heat capacity then results in the overall temperature dependence of
the form
\begin{equation}
  \label{eq:46}
  \ln\left(\tau/ \tau_0\right) = \frac{DT'T}{ T^2 - z\lambda T' }.
\end{equation}
This type of the temperature law was obtained for hard-sphere fluids
by Jagla\cite{Jagla:99} who combined the Adam-Gibbs formalism with an
empirical equation of state. Note, however, that the configurational
entropy in Jagla's model has a $1/T^2$ temperature scaling
inconsistent with the empirical $1/T$ law [Eqs.\ (\ref{eq:4}) and
(\ref{eq:19})].

\begin{widetext}
\begin{table*}[htbp]
  \centering
  \caption{{\label{tab:2}}Best-fit parameters of Eq.\ (\ref{eq:44}) to experimental 
    dielectric relaxation data with $\tau_0$, $D$, and $T'$ considered as fitting parameters. The experimental (superscript ``exp'') and calculate (superscript ``calc'') 
    pressure variation of the glass transition temperature 
    $dT_g/dP$ is given in K/MPa, all temperatures are in K. }
\begin{ruledtabular}
\begin{tabular}{lccccccc}
Substance & $\log (\tau_0/s)$   &   $D$   &  $T'$ & $T'$\footnotemark[1] & $T'$\footnotemark[2]
                       & $dT_g/dP^{\text{exp}}$ &  $dT_g/dP^{\text{calc}}$\\
\hline
Toluene                         &  $-15.6$ & 145  & 9.3   & 9.8 & 8.7 & & \\
$o$-terphenyl (OTP)\footnotemark[3]             
                                & $-14.0$  & 161  & 13.8 & 12.2 & 13.7 & 
                                                                0.26\footnotemark[4] & 0.22\\
2-methyltetrahydrofurane (MTHF) & $-13.5$  & 95.6 & 7.9   & 5.9 & 8.9  &  & \\
Salol                           & $-14.4$  & 204  & 11.9  & 10.5 & 13.3 & 0.20\footnotemark[5] & 0.26\\
3-bromopentane (3BP)            & $-13.6$  & 209  & 7.1   & & 9.4 & & \\
Glycerol                        & $-14.2$  & 159  & 13.2  & & 10.0 & 0.03\footnotemark[6] & 
                                                                   0.05\footnotemark[7] \\
$n$-propanol (nPOH)               & $-13.2$  & 218  & 7.9   & & 6.5  & 0.07\footnotemark[8] & 
                                                                   0.05\footnotemark[9]\\
\end{tabular}
\end{ruledtabular}  
\footnotetext[1]{Calculated for fragile liquids from Eq.\ (\ref{eq:50}) by applying the
  thermodynamic fitting parameters from Table \ref{tab:1}.}
\footnotetext[2]{Based on the steepness fragility index according to Eq.\ (\ref{eq:52}).}
\footnotetext[3]{Fit to Eq.\ (\ref{eq:44}) for OTP was obtained 
  by restricting $\log(\tau_0)$ to be equal to $-14$.}
\footnotetext[4]{From Ref.\ \onlinecite{Atake:79}. }
\footnotetext[5]{From Ref.\ \onlinecite{Casalini:03}. The calculated $dT_g/dP$ refers to $P=300$ MPa instead of atmospheric pressure in case of OTP since the data in Ref.\ 
\onlinecite{Casalini:03} apply to high pressures only. }
\footnotetext[6]{From Ref.\ \onlinecite{Roland:05}.   }
\footnotetext[7]{Using high-temperature $T,P$-data for viscosity from Ref.\ \onlinecite{Cook:94}. }
\footnotetext[8]{From Ref.\ \onlinecite{Takahara:94}. }
\footnotetext[9]{Using the dielectric data at 0.1 MPa and 100 Mpa from Ref.\ \onlinecite{Gilchrist:57}.}
\end{table*}
\end{widetext}

\subsection{Application to experimental data}
\label{sec:4-2}
Equations (\ref{eq:44}) and (\ref{eq:46}) suggest some qualitative
results consistent with experimental observations. First, the model
establishes a direct link between fragility and configurational/excess
heat capacity.  The configurational heat capacity in the denominator
of Eq.\ (\ref{eq:44}) decreases at high temperatures, e.g.\ above the
melting temperature $T_m$.  Therefore, the relaxation kinetics will
change from super-Arrhenius at low temperatures to Arrhenius at high
temperatures.\cite{Frenkel:55} 
The extent of fragile behavior is controlled by $\lambda$ in the denominator
of Eq.\ (\ref{eq:46}) which, according to our thermodynamic analysis,
is large for fragile liquids, resulting in curved Arrhenius plots.
Note that parameter $D$ in Eqs.\ (\ref{eq:44}) and (\ref{eq:46}) is
essentially constant across all liquids studied (Table \ref{tab:2}) in
contrast to parameter $D$ in the VFT equation [Eq.\ (\ref{eq:1})]
which correlates with fragility.  Finally, Eq.\ (\ref{eq:44}) predicts
a return to the Arrhenius behavior\cite{AngellJAP:00} on passing below
$T_g$ because of the drop of $C_P^c$ which defines $T_g$. This feature
has previously been unique to the Adam-Gibbs equation.

\begin{figure}
  \centering
  \includegraphics*[width=6cm]{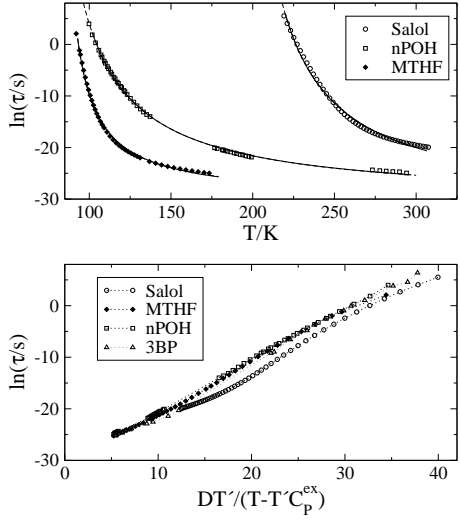}
  \caption{Dielectric relaxation time vs $T$ (upper panel) and vs
    $DT'/(T-C_P^{\text{ex}}T')$ (lower panel) with $D$ and $T'$ listed
    in Table \ref{tab:2}. In the upper panel, the solid lines are fits
    to Eq.\ (\ref{eq:44}) and the dashed lines, indistinguishable from
    the solid lines on the scale of the plot, are the VFT fits.  The
    intercept gives the physical value of $10^{-13}$ s.  }
  \label{fig:11}
\end{figure}

\begin{figure}[htbp]
  \centering
  \includegraphics*[width=6cm]{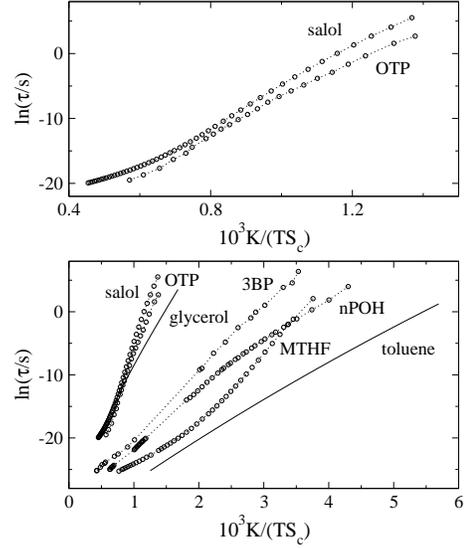}
  \caption{Adam-Gibbs plot of dielectric relaxation time vs
    $(TS_c(T))^{-1}$.  The configurational entropy is calculated from
    the 1G model with the parameters listed in Table \ref{tab:1}.  The
    points are experimental dielectric data from Ref.\
    \onlinecite{Richert:98}, the solid lines are obtained by using
    VFT fits of the experimental data from Refs.\
    \onlinecite{Doss:97} and \onlinecite{Menon:92}. }
  \label{fig:12}
\end{figure}

In addition to providing a qualitatively correct picture, Eq.
(\ref{eq:44}) performs surprisingly well in fitting the dielectric
relaxation times\cite{Richert:98,Doss:97,Menon:92} of both
intermediate and fragile liquids (Fig.\ \ref{fig:11} and Table
\ref{tab:2}), as well as of simulation data (see below). Fits of
experimental dielectric relaxation times to Eq.\ (\ref{eq:44}), with
$\tau_0$, $D$, and $T'$ considered as fitting parameters, are
indistinguishable from VFT fits in all cases studied (Fig.\
\ref{fig:11}, upper panel). $\ln(\tau)$ plotted against
$DT'/(T-T'C_P^{\text{ex}})$ also yields straight lines and physically
reasonable intercepts (Fig.\ \ref{fig:11}, lower panel). The quality
of the linear correlations is as good as for the Adam-Gibbs plot (cf.\
Figs.\ \ref{fig:11} and \ref{fig:12}) and in some cases is even better
(MTHF). In all fits, experimental $C_P^{\text{ex}}(T)$ were used
instead of $C_P^c(T)$ suggested by the denominator of Eq.\
(\ref{eq:44}). In the 1G model, $C_P^{\text{ex}} \simeq C_P^c$ for fragile
liquids.

The nominator parameter $D$ in Eq.\ (\ref{eq:44}) can be related to
the average number of excitations in the hot region via Eq.\
(\ref{eq:45}). From Table \ref{tab:2}, we obtain $N_c\simeq 10-20$. With
the usual molecular diameter of $\sigma\simeq 5.5$ \AA, this number projects into
a length-scale of $\simeq 1.5$ nm for a spherical cluster or larger if
relaxation is facilitated through chains of
molecules.\cite{Gebremichael:05} This length is in general accord with
current estimates of spatial heterogeneities in supercooled
liquids.\cite{Donth:91} The activation barrier scales quadratically
with the number of units in the hot region compared to the linear
scaling with the size of the cooperatively rearranging region of the
Adam-Gibbs theory. However, there is no divergent length-scale in the
present formulation.

\begin{figure}
  \centering
  \includegraphics*[width=6cm]{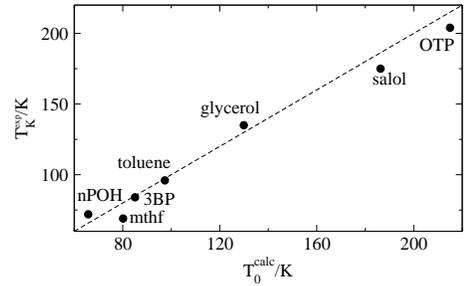}
  \caption{{\label{fig:13}}Experimental Kauzmann temperature $T_K^{\text{exp}}$ 
          plotted against the temperature $T_0^{\text{calc}}$ at which 
          the relaxation time in Eq.\ (\ref{eq:44}) diverges. This temperature
          is calculated by solving Eq.\ (\ref{eq:49}) in which temperature
          $T'$ is taken from Table \ref{tab:2}. }  
\end{figure}

Activated dynamics considered here do not anticipate the divergence of
the relaxation time at any finite temperature, and the divergent
solution given by Eqs.\ (\ref{eq:44}) and (\ref{eq:46}) is an artifact
of the approximate integration used in Eq.\ (\ref{eq:42}). However,
the mathematical solution of the equation
\begin{equation}
  \label{eq:49}
  T_0 = T' C_P^c(T_0)
\end{equation}
can be associated with the VFT temperature $T_0$ often
reported experimentally. The temperature $T_0$ calculated from this
equation (Fig.\ \ref{fig:13}) is almost equal to the experimental Kauzmann
temperature $T_K^{\text{exp}}$ for all liquids listed in Table
\ref{tab:2}. Equation (\ref{eq:44}) is therefore consistent with the
empirically documented accord between $T_0$ and
$T_K^{\text{exp}}$.\cite{Angell:97}

For fragile liquids, Eq.\ (\ref{eq:49}) simplifies to [Eq.\
(\ref{eq:16-1}), $C_P^{\text{ex}}\simeq C_P^c$]
\begin{equation}
  \label{eq:50}
  T' = \frac{T_0^2}{T_g C_P^c(T_g)} = \frac{\lambda}{zs_0^2}. 
\end{equation}
The temperature $T'$ calculated  from Eq.\
(\ref{eq:50}), with the thermodynamic parameters from Table
\ref{tab:1}, is indeed close to the direct fit of Eq.\ (\ref{eq:44})
to experimental relaxation data (Table \ref{tab:2}).  The parameter
$T'$ can also be related to the steepness fragility index [Eq.\
(\ref{eq:20})].  Assuming $T^{-1}$ scaling for $C_P^c(T)$ in Eq.\
(\ref{eq:44}), one gets
\begin{equation}
  \label{eq:52}
  T' = \frac{T_g}{C_P^c(T_g)} \frac{m - m_{\text{min}}}{m+m_{\text{min}}},
\end{equation}
where $m_{\text{min}}=\log(\tau(T_g)/ \tau_0)\simeq 16 $. Estimates of $T'$ based
on this equation are also listed in Table \ref{tab:2}. Since, for
fragile liquids, $T'$ is related to observable quantities through Eq.\
(\ref{eq:50}), one can derive the following relation for the kinetic
fragility
\begin{equation}
  \label{eq:52-1}
  \frac{m}{m_{\text{min}}} = \frac{1+ (T_0/T_g)^2}{1-(T_0/T_g)^2} .
\end{equation}
This equation, which was previously derived by Ruocco \textit{et
  al}.,\cite{Ruocco:04} holds reasonably well for fragile liquids
($m>53$).  It makes a connection between the steepness index and the
ratio $T_0/T_g$ recommended by some authors as a measure of
fragility.\cite{Donth:01}

Since the configurational heat capacity of fragile liquids 
is linearly related to the configurational
entropy, $s_c = s_0^c - c_P^c(T)$, Eq.\ (\ref{eq:44})
can be re-written in terms of the configurational entropy
\begin{equation}
  \label{eq:53}
  \ln(\tau/ \tau_0) = \frac{DT'}{T + T'(S_c - zs_0^c)} .
\end{equation}
The current formulation still provides a link between the relaxation
time and configurational entropy, but the algebra is different from the
Adam-Gibbs relation [Eq.\ (\ref{eq:2})].  The validity of either
functional dependence of the relaxation time is often tested by
calculating the slope of the glass transition temperature with
pressure.  Equation (\ref{eq:44}) suggests that $T/T' - C_P^c(T)$ remains
constant at the glass transition temperatures measured at different
pressures. Following Goldstein,\cite{Goldstein:63,DebenedettiBook:96}
the condition $d(T/T' - C_P^c(T))=0$ provides the slope of the glass
transition temperature with pressure, $dT_g/dP$.  One needs to
calculate the derivative of $T'$ over pressure and the pressure and
temperature derivatives of $C_P^c$.  The pressure derivative gives $(\partial
C_P^c/ \partial P)_T=- T_g V_g \Delta\alpha_P^2/ k_{\text{B}}$, where $V_g$ is the
molecular volume of the glass.  The difference of the isobaric
expansivities of the liquid and glass, $\Delta \alpha_P$, is in the range $\simeq 5\times
10^{-4}$ K$^{-1}$ (Refs.\ \onlinecite{Goldstein:63,Atake:79}) allowing
one to neglect the pressure derivative of $C_P^c$. One then gets
\begin{equation}
  \label{eq:54}
  \frac{dT_g}{dP} = \frac{T_g}{T'}\,\frac{\partial T'}{\partial P} \left(1-T'(\partial C_P^c/ \partial T)_P \right)^{-1} .
\end{equation}
Equation (\ref{eq:54}) should be compared to what follows from the
Adam-Gibbs scaling\cite{DebenedettiBook:96}
\begin{equation}
  \label{eq:55}
  dT_g/dP = T_gV_g\Delta \alpha_p / (S_c + C_P^c) .
\end{equation}

Most data, available for strong/intermediate
liquids,\cite{Goldstein:63,Angell:76,Takahara:94} indicate that Eq.\ 
(\ref{eq:55}) adequately describes the experiment. Since $\partial T_0/ \partial
P>0$ (Refs.\ \onlinecite{Takahara:94,Cook:94,Casalini:01}), one can
expect from Eq.\ (\ref{eq:50}) that $\partial T' /\partial P>0$. The actual numbers
for this derivative can be obtained from $T,P$ relaxation
data.\cite{Gilchrist:57,Atake:79,Cook:94,Takahara:94,Dreyfus:03,Casalini:03,Roland:05}
The results of these calculation are given in Table \ref{tab:2}. The
relaxation times were fitted to Eq.\ (\ref{eq:44}) at ambient pressure
with $D$ and $T'$ considered as fitting parameters and experimental
$C_P^{\text{ex}}(T)$ and ambient pressure. The parameter $D$ was then kept
constant at elevated pressures and $\partial T' /\partial P$ was evaluated from the
fit. The results of these calculations are in reasonable agreement
with reported $dT_g/dP$ and reproduce the drop of this derivative in
going from fragile to intermediate glass-formers (Table \ref{tab:2}).

\begin{figure}
  \centering
  \includegraphics*[width=6cm]{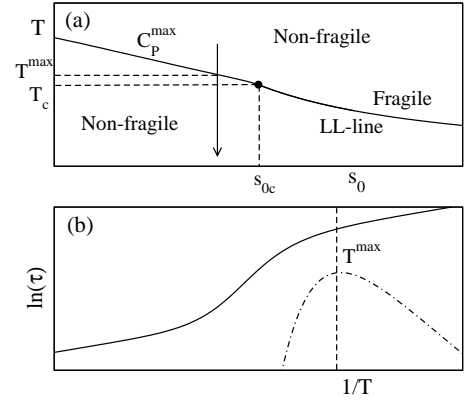}
  \caption{The liquid-liquid coexistence line (``LL-line'') and the
    line of maximum heat capacity (``Widom line''\cite{Xu:05}) in
    the $(T,s_0)$ parameters plane (a).  Thermodynamic fragile
    behavior is limited by the critical point $(T_c,s_{0c})$ [Eq.\
    (\ref{eq:22})]. The vertical arrow shows the cooling path along
    which the relaxation time is calculated using Eq.\ (\ref{eq:44})
    (b).  $T^{\text{max}}$ denotes the temperature at which the of
    heat capacity (dash-dotted line in (b)) passes through a maximum,
    $C_P^{\text{max}} = C_P(T^{\text{max}})$. All the plots have been
    generated at constant $\epsilon_0$, $\lambda$ and $z$. }
  \label{fig:111}
\end{figure}

\section{Discussion}
\label{sec:5}
The model developed here describes the thermodynamic properties of
glass-formers in terms of statistics of excitations from a
single-energy state of the ideal glass to a Gaussian manifold of
configurationally unjammed states with higher energy and entropy. The
model suggests that the thermodynamic signatures of fragile liquids,
in particular a sharply increasing configurational heat capacity close
to the glass transition, can be identified with the existence of a
thermodynamic phase transition (``LL-line'' in Fig.\ \ref{fig:111}a),
usually hidden below the glass transition
temperature.\cite{Gibbs:58,Sethna:91,Klein:94} The strong/fragile
behavior is distinguished within the model by two parameters, the
trapping energy parameter $\lambda$ and the entropy gain for a single
configurational excitation $s_0$, the ratio of them equal to the
experimental Kauzmann temperature [Eq.\ (\ref{eq:16})]. Both
parameters significantly decrease when going from fragile to
intermediate/strong liquids, with the parameter $\lambda$ showing the most
significant change (Table \ref{tab:1}). Fragile liquids are therefore
characterized by a broad range of configurationally excited states,
whereas the energy distribution for intermediate/strong liquids is
very narrow (Fig.\ \ref{fig:1}).  Because of low values of either $\lambda$
or $s_0$ (or both of them) strong/intermediate liquids fall in the
range of temperatures and excitation entropies [$T>T_c$, $s_0<
s_{0c}$, Eq.\ (\ref{eq:22})] not allowing a first-order transition.

\begin{figure}[h]
  \centering
  \includegraphics*[width=6cm]{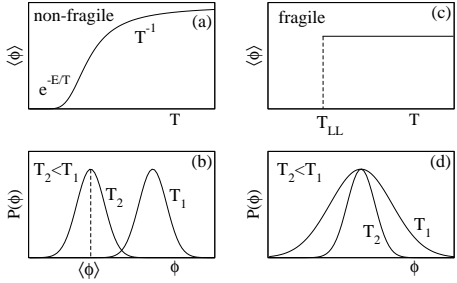}
  \caption{Illustration of the temperature behavior of the average
    basin energy $\langle \phi \rangle$ and the distribution of basin energies $P(\phi)$
    in non-fragile and fragile liquids. The width of the basin energy
    distribution scales as $1/ \sqrt{N}$ and requires small systems to
    be observed in simulations. }
  \label{fig:14}
\end{figure}

\begin{figure}
  \centering
  \includegraphics*[width=6cm]{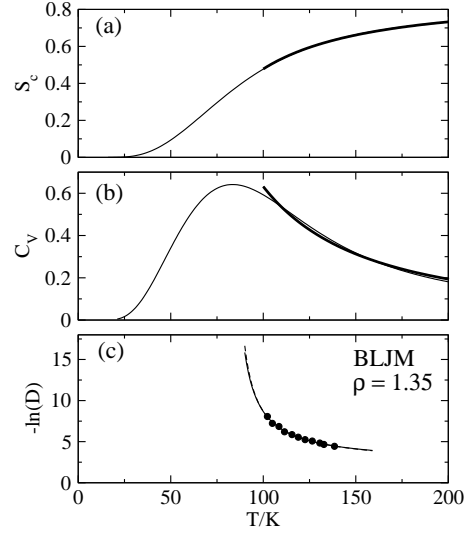}
  \caption{Configurational entropy (a), configurational
    (constant-volume) heat capacity (b), and diffusion coefficient (c)
    of the 80(A):20(B) BLJM liquid at $\rho\sigma_{AA}^3=1.35$, where $\sigma_{AA}$
    is the LJ diameter of the component with higher concentration.
    The thick solid lines are the results of simulations from Ref.\
    \onlinecite{Sastry:01} arbitrarily limited to $T>100$ K.  The
    diffusion coefficients from Ref.\ \onlinecite{Sastry:00} (circles,
    c) are fitted to Eq.\ (\ref{eq:44}) (solid line) and to the
    VFT relation (dashed line). The two fits are almost
    indistinguishable on the scale of the plot.  The fitting to 1G
    model [Eqs.\ (\ref{eq:11}), (\ref{eq:14-1}), and (\ref{eq:44})]
    results in the following fitting parameters: $\epsilon_0=219$ K, $\lambda=71$
    K, $s_0^c =0.32$, $T'=92.8$ K, and $D=3.1$.  The
    thermodynamic Kauzmann temperature $T_K$ is zero when calculated
    in the 1G model (the thin lines in a and b). }
  \label{fig:15}
\end{figure}

The critical point $(T_c,s_{0c})$, separating fragile from non-fragile
liquids (Fig.\ \ref{fig:111}a), makes the descent into the energy
landscape qualitatively different for them. The average basin energy
of non-fragile liquids first starts to drop from the high-energy
plateau according to the $1/T$ law and then inflects into an
exponential decay.  The distribution of basin energies is narrow, and
its maximum shifts to lower energies with cooling (Fig.\
\ref{fig:14}a,b).  This behavior, often observed in simulations of binary
LJ liquids,\cite{Sastry:98,Buchner:99,Mossa:02,Gebremichael:05} is
well reproduced by the present model which places these fluids into
the strong/intermediate category. As an example, we show in Fig.\
\ref{fig:15} the fit of the 1G model to Sastry's data on binary LJ
mixture (BLJM).\cite{Sastry:00,Sastry:01} The calculations were done
for the highest density studied in simulations at which the BLJM
liquid is most fragile.  Equations (\ref{eq:11}) and (\ref{eq:14-1})
were used, respectively, for the configurational entropy and heat
capacity, while Eq.\ (\ref{eq:44}) was applied to the simulated diffusion
coefficient. The excitation energy ($\epsilon_0=219$ K) and the trapping
energy ($\lambda=71$ K) obtained from the fit produce the following critical
parameters: $T_c=35.5$ K and $s_{0c}=4.2$ (accidentally, $T_c$ is
close to $T_K^{\text{exp}}$).  On the other hand, the excitation
entropy from the fit, $s_0^c=0.32$, and the temperature range $T>T_c$
together put the BLJM liquid into the category of intermediate
fragility.  With such low excitation entropy, the population of the
excited state is close to 0.5 at high temperatures. From Eq.\ (\ref{eq:11}) 
one then gets ($z=1$) $S_0=s_0^c/2+ \ln(2) \simeq 0.85$. The same number, 0.85, is
reported by Sastry for the top of the enumeration function
$s^c_{\text{max}}$.\cite{Sastry:01}

The behavior of fragile liquids is quite different. The
configurationally excited state is almost entirely populated ($x\simeq 1$)
in the entire range of experimentally accessible temperatures down to
the glass transition. The average basin energy is almost
temperature-independent, and the only effect of lowering the
temperature on the distribution of basin energies is its narrowing
according to the fluctuation-dissipation theorem (Fig.\ \ref{fig:14}c,d).
The invariant population of the configurationally excited states is
abruptly terminated at the equilibrium first-order transition or
(on supercooling) at spinodal instability at which point it drops to a low value
representative of an entropy-poor, low-temperature phase as experimentally
confirmed for TPP.\cite{Kurita:04}

\begin{figure}[htbp]
  \centering
  \includegraphics*[width=6cm]{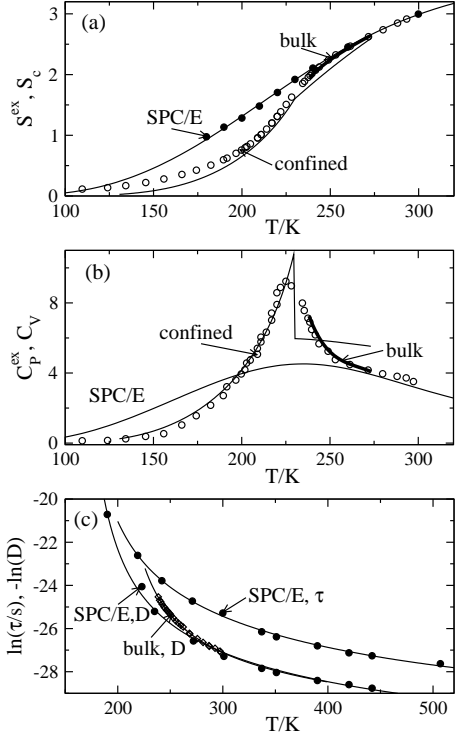}
  \caption{{\label{fig:16}} (a) Excess entropy for bulk laboratory
    water (``bulk''), nanoconfined water (``confined''), and
    configurational entropy for SPC/E water (``SPC/E''). The thick
    line and open points refer to bulk\cite{AngellW:80} and
    nanoconfined experiments, respectively. The closed points refer to
    constant-volume simulations.\cite{Starr:03} (b) Experimental
    excess heat capacity for bulk (thick line\cite{AngellW:80}) and
    hard nanoconfined\cite{Maruyama:04} (open points) water.  The thin
    lines in (a) and (b) are fits to the thermodynamic 1G model [Eqs.\
    (\ref{eq:11-1})--(\ref{eq:12})].  The configurational heat
    capacity of SPC/E water in (b) is calculated from the 1G model
    based on the fit of the configurational entropy in (a). (c)
    Simulation (constant-volume) results for the Debye relaxation time
    (closed points) and diffusivity (open points) of SPC/E
    water.\cite{DMjpcb1:06} Diamonds show experimental diffusivity of
    laboratory water.\cite{Price:99} The solid lines are fits to Eq.\
    (\ref{eq:44}) with the configurational heat capacity obtained by
    numerical differentiation of $S_c$ for SPC/E water (see Table
    \ref{tab:3} for the fitting parameters).  The diffusivity curves
    have been arbitrarily shifted to fit to the scale of the plot. For
    all three panels, the closed points refer to simulations and open
    points are used to indicate the laboratory experiment. }
\end{figure}
 
The fit of the model to experimental excess heat capacities and
entropies of molecular glass-formers (Table \ref{tab:1}) has pushed
the liquid-liquid phase transition below the glass transition.  For
some substances,\cite{AngellJAP:00,Tanaka:00,Starr:03,Xu:05} the
thermodynamic liquid-liquid transition can be found above the glass
transition.  As an example of this situation, the 1G model is applied
to excess entropy and heat capacity of laboratory
water.\cite{AngellW:80,Maruyama:04} The thick lines in Fig.\
\ref{fig:16}a,b show the results of measurements on bulk
samples,\cite{AngellW:80} while open circles 
refer to samples confined in 3 nm pores of silica
gel.\cite{Maruyama:04} The two sets of data coincide at high
temperatures. Because of the limited temperature range of the bulk
water data the fit to the thermodynamic 1G model is done for
nanoconfined water (Table \ref{tab:3}). The 1G model predicts a weak
first-order phase transition at $T_{LL}=229$ K ($P=1$ atm).  This
number is very close to previous estimates for the temperature of
crossing the ``Widom line'' at which the heat capacity gets a
maximum\cite{Starr:03} and more recent measurements placing this point
at 223 K.\cite{Mallamace:06} After the kink at $T_{LL}$, the entropy
slowly decays to zero at $T_K \to 0$.

The behavior of excess thermodynamic parameters of the laboratory
water is compared to configurational parameters of SPC/E water
obtained from constant-volume simulations of Starr \textit{et
  al.}\cite{Starr:01,Starr:03} The fit of Eqs.\
(\ref{eq:11})--(\ref{eq:12}) to the configurational entropy of SPC/E
water (Fig.\ \ref{fig:16}a and Table \ref{tab:3}) yields a low
excitation entropy $s_0^c=1.3$, below the critical value of
$s_{0c}=4.0$ [Eq.\ (\ref{eq:22})].  The heat capacity curve then shows
a broad pre-critical peak reminiscent of, but broader than, the heat
capacity of nanoconfined water (Fig.\ \ref{fig:16}b). As in the case
of the BLJM liquid, the excited state population, characterizing the
configurational manifold, is slightly above 0.5 at high temperatures
resulting in the plateau of the configurational entropy at ($z=3$)
$S_0 \simeq z(s_0^c/2+\ln(2))\simeq 4$. This number is close to the estimate for
the top of the enumeration function given by
Sciortino,\cite{Sciortino:05} $S_0\simeq 3.7$ at $P\simeq 0$.

The comparison of fits to laboratory and simulation data shows that
the configurational part of the excitation entropy $s_0^c$ makes 
about 46\% of $s_0$ for SPC/E water and only about 30\% for $o$-terphenyl
(Fig.\ \ref{fig:6}).  These comparable fractions of configurational
components suggest that it is the change in the vibrational density of
states (around the boson peak at $\simeq 30$ cm$^{-1}$) caused by
configurational excitations that is primarily responsible for the
fragile behavior of glass-formers.\cite{Angell:02}

The configurational entropy from simulations can next be used to fit
the relaxation times by applying Eq.\ (\ref{eq:44}).  This procedure
is straightforward for the Debye relaxation time and diffusivity
obtained from constant-volume simulations\cite{DMjpcb1:06} at the same
conditions as the data shown in Fig.\ \ref{fig:16}a (see Table
\ref{tab:3} for the fitting parameters). The application to laboratory
data is less obvious because of possible differences between
configurational heat capacities at constant volume and constant
pressure. Nevertheless, diffusivity of laboratory water\cite{Price:99}
can be fitted to Eq.\ (\ref{eq:44}), although with the fitting
parameters (Table \ref{tab:3}) requiring more fragile behavior (cf.\
triangles to open circles in Fig.\ \ref{fig:16}c).

\begin{table}
  \centering
  \caption{Fit of the 1G model to excess thermodynamics\cite{AngellW:80,Maruyama:04} 
    and diffusivity\cite{Price:99}
    of laboratory water and to configurational entropy, diffusivity, and
    Debye relaxation time of SPC/E water from constant-volume 
    simulations\cite{Starr:01,Starr:03,DMjpcb1:06} ($\rho=1.0$ g/cm$^3$). 
    The heat capacity and excess entropy of nanoconfined water\cite{Maruyama:04} 
    are used
    for the thermodynamic fitting. The 
    energy parameters are in K, $D$ and $z$ are dimensionless. }
\begin{ruledtabular}
  \begin{tabular}{lcccccccc}
Liquid & $\epsilon_0$ & $\lambda$ &  $s_0$ & $s_{0c}$ & $z$ & $D$ & $T'$ & $\log(\tau_0/s)$ \\    
\hline
Bulk/Confined water  & 1298 & 649 & 2.8  & 2.0 & 3.0 & 66\footnotemark[1] & 19 &  \\
SPC/E water & 751  & 251 & 1.3\footnotemark[2]  & 4.0 & 3.0 & 282\footnotemark[3] 
                                                & 5.7 & $-13.4$\\
            &      &     &                      &     &     & 193\footnotemark[4] 
                                                            & 7.6 & \\
  \end{tabular}
\end{ruledtabular}
\footnotetext[1]{ From fitting the diffusivity of laboratory water according
                  to Price \textit{et al.}\cite{Price:99} }
\footnotetext[2]{ Configurational component $s_{0}^c$ of the excitation entropy. } 
\footnotetext[3]{ From fitting the Debye relaxation time.}
\footnotetext[4]{ From fitting the diffusivity. }
  \label{tab:3}
\end{table}

Equation (\ref{eq:44}) is the central result of our modeling of
relaxation. It is also consistent with the empirically documented
accord between the VFT and Kauzmann temperatures (Fig.\ \ref{fig:13}).
The model combines the thermodynamics of configurational excitations
with the notion that assembling excited units in clusters is required
for activated relaxation.\cite{Douglas:06} The thermodynamics of
glass-formers is thus determined by the statistics of excitations,
while dynamics probe more rare events of clustering of excitations.
One of the signatures of thermodynamic driving force behind relaxation
is the realization of a dynamic fragile-to-strong transition close to
the point where the heat capacity in the denominator of Eq.\
(\ref{eq:44}) reaches its maximum, $C_P^{\text{max}}$. This
sub-critical maximum is indeed predicted by the thermodynamic 1G model
is the vicinity of the critical point $(T_c,s_{0c})$ (Fig.\
\ref{fig:111}a). When the cooling path goes close to the critical
point, the fragile non-Arrhenius dynamics changes to strong Arrhenius
dynamics after passing the line of the heat capacity maximum (``Widom
line'',\cite{Xu:05} Fig.\ \ref{fig:111}b). A similar behavior has
recently been observed for nanoconfined
water.\cite{Liu:05,Mallamace:06}

Experimental evidence supports a link between the viscosity and
configurational
thermodynamics\cite{Scherer:90,Sastry:01,Gebremichael:05,Saika-Voivod:06}
but this does not necessarily mean that activated events are driven by
the entropy. Equation (\ref{eq:44}) still connects the relaxation time
to configurational entropy [see Eq.\ (\ref{eq:53})], but is based on
standard arguments of activated kinetics in terms of excess kinetic
energy accumulated at a ``mobile unit''.  The model thus puts focus on
the kinetic energy and its fluctuations as the main driving force of
liquid relaxation.\cite{Ferrer:98} This notion brings relaxation of
supercooled liquid in general accord with approaches developed to
describe activated events in chemistry where accumulation of
sufficient kinetic energy along a reaction coordinate, and not the
entropy, is viewed as the general mechanism behind activated
transitions. Since relaxation is understood in terms of clusters of
excitations, the model can potentially bridge to more macroscopic
arguments of elastic theories considering a region of shear
displacements shoving the environment from a relaxing
unit.\cite{Dyre:06}

The present model does not anticipate a divergent length-scale and
also avoids the need to consider the mosaic interface energies of
Ref.\ \onlinecite{Xia:00}, which recent simulations have failed to
support.\cite{Cavagna:06} It remains to be seen whether the formalism
of configurational
excitations,\cite{Goldstein:72,Angell:72,Garrahan:03,Tanaka:03,DMjcp5:05}
``dressed'' with the relevant excitation thermodynamics, provides a
sufficient basis for the theoretical modeling of supercooled liquids.
While the model yields a set of simple equations for some basic
properties commonly reported for glass-formers [e.g., Eqs.\
(\ref{eq:16-1}), (\ref{eq:16}), (\ref{eq:22}), and (\ref{eq:44})], it
should be recognized that the mean-field character of the model may
push it too much into a strong first-order type of thermodynamic
transition.  Some ``softening'' of the model including fluctuations
around the mean field would result in greater flexibility, but this
would only come at the expense of an increase in the number of
parameters.

\begin{acknowledgments}
  The authors are grateful to Ranko Richert for the help with
  dielectric relaxation data and to Srikanth Sastry for useful
  discussions.  This work was supported by the NSF through the grants
  CHE-0616646 (D.\ V.\ M.) and DMR-0082535 (C.\ A.\ A.). C.\ A.\ A.\
  thanks Francesco Sciortino for a number of illuminating discussions. 
  D.\ V.\ M.\ thanks Karl Freed for critical comments on the manuscript. 
\end{acknowledgments}

\bibliographystyle{apsrev}
\bibliography{/home/dmitry/p/bib/chem_abbr,/home/dmitry/p/bib/photosynth,/home/dmitry/p/bib/liquids,/home/dmitry/p/bib/glass,/home/dmitry/p/bib/et,/home/dmitry/p/bib/dm,/home/dmitry/p/bib/dynamics,/home/dmitry/p/bib/ferro}

\end{document}